\documentclass{aa}
\usepackage{booktabs}
\usepackage{lscape} 
\usepackage{caption}
\usepackage{color, colortbl}
\usepackage{hyperref}
\usepackage[online]{threeparttablex}
\hypersetup{
    colorlinks=true,
    linkcolor=blue,
    citecolor=blue,
    filecolor=black,      
    urlcolor=blue,
    }
\usepackage[dvipsnames]{xcolor}

\usepackage{graphicx}
\usepackage{txfonts}
\usepackage{amsmath}
\usepackage{mathtools}
\usepackage{multirow}
\usepackage{longtable}
\usepackage{todonotes}
\usepackage{version}
\usepackage{natbib}
\usepackage[switch]{lineno}
\usepackage{comment}

\begin{document} 

\title{CHEX-MATE: X-ray surface brightness discontinuities across a representative cluster sample}

 \author{M. G. Campitiello\inst{1} \fnmsep\thanks{\email{mcampitiello@anl.gov}}
          \and
          S. Ettori\inst{2,3}
          \and
          L. Lovisari\inst{4,5}
          \and F. Gastaldello\inst{4}
          \and H. Bourdin\inst{6,7} 
          \and P. Mazzotta\inst{6,7}
          \and M. Rossetti\inst{4}
          \and M. Balboni\inst{4,8}
          \and A. Botteon\inst{9} 
          \and R. Cassano\inst{9}
          \and F. De Luca\inst{6,7} 
          \and D. Eckert\inst{10}
          \and W. Forman\inst{5} 
          \and M. Gaspari\inst{11} 
          \and S. Ghizzardi\inst{4}
          \and M. Gitti\inst{8,9}
          \and S. T. Kay\inst{12}
          \and B. J. Maughan\inst{13} 
          \and E. Pointecouteau\inst{14} 
          \and G. W. Pratt\inst{15}
          \and E. Rasia\inst{16,17} 
          \and J. Sayers\inst{18}
          \and M. Sereno\inst{2,3}}

   \institute{High Energy Physics Division, Argonne National Laboratory, Lemont, IL 60439, USA 
    \and   
   INAF, Osservatorio di Astrofisica e Scienza dello Spazio, via Piero Gobetti 93/3, 40129 Bologna, Italy
         \and
         INFN, Sezione di Bologna, viale Berti Pichat 6/2, 40127 Bologna, Italy
         \and
         INAF -- IASF Istituto di Astrofisica Spaziale e Fisica Cosmica di Milano, Via Alfonso Corti 12, 20133 Milano, Italy
         \and
        Center for Astrophysics | Harvard \& Smithsonian, 60 Garden St.,
Cambridge, MA 02138, USA,
USA  
        \and
        Dipartimento di Fisica, Università di Roma ‘Tor Vergata’, Via della Ricerca Scientifica 1, 00133 Roma, Italy
        \and
INFN, Sezione di Roma ‘Tor Vergata’, Via della Ricerca Scientifica, 1, 00133 Roma, Italy 
\and Dipartimento di Fisica e Astronomia (DIFA), Universit\`a di Bologna, via Gobetti 93/2, 40129 Bologna, Italy
\and Istituto Nazionale di Astrofisica (INAF) – Istituto di Radioastronomia (IRA), via Gobetti 101, I-40129 Bologna, Italy
\and Department of Astronomy, University of Geneva, Ch. d’Ecogia 16, CH-1290 Versoix, Switzerland
\and Department of Physics, Informatics and Mathematics, University of Modena and Reggio Emilia, 41125 Modena, Italy
\and  Jodrell Bank Centre for Astrophysics, Department of Physics and Astronomy, The University of Manchester, Oxford Road,
Manchester M13 9PL, UK
\and HH Wills Physics Laboratory, University of Bristol, Tyndall Ave, Bristol, BS8 1TL, UK
\and IRAP, CNRS, Université de Toulouse, CNES, UT3-UPS, Toulouse,
France
\and  Université Paris-Saclay, Université Paris Cité, CEA, CNRS, AIM,
91191 Gif-sur-Yvette, France
\and INAF – Osservatorio Astronomico di Trieste, Via Tiepolo 11, 34131
Trieste, Italy
\and
IFPU – Institute for Fundamental Physics of the Universe, Via Beirut
2, 34151 Trieste, Italy
\and California Institute of Technology, 1200 E. California Blvd., MC 367-17, Pasadena, CA 91125, USA
}

\date{Accepted: 31 August 2026}

\abstract{
We analyze the presence of X-ray surface brightness (SB) discontinuities, such as shocks and cold fronts, in the intracluster medium (ICM) of the 116 galaxy clusters from the CHEX-MATE sample. These features arise from accretion-related processes, including the merging of subclumps onto the main cluster halo and the sloshing of cold gas in a higher-entropy environment. We identify these structures in the XMM CHEX-MATE maps by examining both the SB residuals, obtained by subtracting a cluster model from the observations, and the gradient variations across X-ray images. We validate our method with an extensive analysis and comparison of the discontinuities described already in the literature. Our analysis reveals 66 discontinuities, 32 of which are newly detected, in 48 objects ($\sim$41\%). We find that discontinuities in relaxed systems tend to be closer to the core, weaker, and less aligned with the global X-ray morphology compared to those in disturbed systems. Using wavelet-based temperature maps, we provide a preliminary classification of these discontinuities,
proposing the occurrence of 8 putative shocks and 34 cold fronts. 
Furthermore, about 48\% of the systems with extended radio emission show evidence of ICM discontinuities; conversely, $\sim$88\% of the objects with detected discontinuities have an associated diffuse radio emission.
This work also explores the limitations of using XMM data, characterized by relatively low spatial resolution, for detecting and analyzing such features, highlighting the challenges and potential biases introduced by instrumental constraints.
}


\maketitle
\nolinenumbers

\section{Introduction}

Galaxy clusters are the largest virialized systems in the Universe and their total mass is dominated by dark matter  (DM, $\sim$ 80 \%). During structure formation, this poorly understood component forms deep potential wells, where baryonic matter flows and virializes. The gravitational energy at play is dissipated by shocks and large-scale turbulent motions and goes mainly into heating the majority of the baryons, giving rise to a hot ($kT \sim 2-10$ keV) and tenuous ($n \sim 10^{-3}-10^{-4}$ cm$^{-3}$) plasma, the intracluster medium. The ICM emits mainly via bremsstrahlung in the X-ray band and the study of its distribution and of its properties is crucial to constraining the cluster potential and, therefore, the cluster total mass. 

Although early X-ray observations suggested that the ICM surface brightness could be described by a smooth $\beta$-model \citep{Cavaliere1976}, departures from this simple distribution were already revealed by ROSAT \citep{Churazov2000,Churazov2001}. With the advent of higher-resolution instruments, such as Chandra and XMM, these deviations could be spatially resolved and characterized in greater detail, revealing sharp surface-brightness discontinuities and substructures associated with the dynamical evolution of the ICM \citep[e.g.][]{Markevitch2000,Markevitch2001}. In this work, we adopt the following terminology:
substructures are bound or quasi-bound overdensities, that are typically associated with infalling subhaloes; discontinuities or edges are abrupt changes in the surface-brightness gradient across a narrow radial range, identified in images or profiles and correspond to discontinuities in gas properties. 
The latter features include shocks, where density, temperature, and pressure rise across the front (Mach number $\mathcal{M}>1$), and cold fronts (contact discontinuities), where density increases while temperature decreases so that pressure remains approximately continuous \citep[for a foundational review]{Markevitch2007}. In galaxy clusters, shocks can be driven by the expansion of an active galactic nucleus (AGN) jet in the ICM (in this case they are expected to be observed in a central region of $\sim 100$ kpc, e.g. \cite{Birzan2004, Nulsen2005, Fabian2006, hlavacek2012, Liu2019,Ubertosi2023}), or by merger events in which the subcluster has sufficient gravitational potential to retain some of its proper baryonic mass \citep[e.g.,][]{Markevitch2007}. The detection of shocks is not trivial, since they rapidly propagate to the low-brightness outskirts and become transonic via simple Sedov expansion \citep[e.g.,][]{Gaspari2011}, even when injected in the core. For completeness, we note that simulations also predict accretion-driven shocks forming at large cluster centric distances \citep[e.g.,][]{Miniati2000}. These types of shocks would be expected at large distances from the cluster core, near the virial radius. For this reason, their detection in X-rays is difficult because of their low surface brightness at large radii. Nevertheless, \cite{Akamatsu2017} discuss the possible detection of this type of shock between the clusters Abell 399 and Abell 401. Similarly \cite{Zhang2020} discussed a possible accretion shock in Perseus, and analyses of the Sunyaev–Zeldovich (SZ) have revealed similar features, e.g. in A2319 \citep{Hurier2019}.

While merger shocks were long expected from simulations, the discovery of cold fronts with Chandra sparked a broad effort to understand their origin and microphysics. Initially, they were detected in disturbed systems, such as A2142 \citep{Markevitch2000}, A3667 \citep{Vikhlinin2001a,Vikhlinin2001b,Vikhlinin2002}, and 1E0657-56 \citep{Markevitch2002}, and have been interpreted as contact discontinuities between a surviving cool core of an infalling subcluster and the hotter ambient intracluster medium \citep{Markevitch2000}. Subsequent detections in otherwise relaxed clusters \citep[e.g. A1795][]{Markevitch2001} broadened this picture. 
Minor mergers can excite long-lived gas sloshing in the cluster potential, producing multiple fronts and redistributing metals \citep{Roediger2011, Roediger2012, zuhone2016}. The observed disposition drawn by sloshing cold fronts depends on projection, with edges giving rise to a spiral when the sloshing plane is near the plane of the sky, or as a sequence of arcs or ripples when there is a substantial line-of-sight component \citep{Roediger2011, Ueda2019}. Further analyses of Chandra and XMM images have shown that cold fronts are common \citep[e.g.,][]{Owers2009, Ghizzardi2010, Botteon2018}, and can provide insights on the ICM physics. For instance, recent deep observations reveal Kelvin–Helmholtz eddies at several fronts, enabling constraints on effective viscosity and conduction and pointing to magnetic draping as a stabilizing mechanism \citep{Lyutikov2006, ZuHone2013b, Ichinohe2017, Wang2018, Li2026}. Furthermore, sloshing is not restricted to the cluster core, but can extend to larger scales \citep[e.g.][]{Rossetti2013, Ichinohe2015, Simionescu2012, Ichinohe2019, Walker2022}. In this context, cold fronts have been detected at large radii in clusters such as Abell 3558, highlighting their presence well beyond the core region \citep{Mirakhor2023}, and detailed analyses in systems like PLCKG287.0+32.9 continue to reveal both cold fronts and shocks in dynamically intermediate systems \citep{Gitti2025}. Theoretical modeling of their formation and stability is also evolving, with recent work connecting cold fronts to global eigenmodes of sloshing gas, even in weakly magnetized environments \citep{Choudhury2025}. 
The study of these edges also provides key insights into the generation and distribution of non-thermal components in the ICM, as they often spatially coincide with radio relics and halos \citep[e.g.,][for a review]{Giacintucci2008, Brown2011, Macario2011, Shimwell2015, Botteon2016, vanweeren2019} or the sloshing spiral pattern described above and the boundaries of radio mini-halo \citep[e.g.][]{Mazzotta2008, Giacintucci2014, Giacintucci2014b}. Recently, high-quality radio observations have also enabled us to detect a spatial coincidence between X-ray edges and SB edges in radio halos \citep{Botteon2023}. All these results demonstrate that the thermal and non-thermal emissions of galaxy clusters are strictly connected, with the latter one being shaped by the dynamical motions that occur in the ICM.

In this paper, we perform a systematic search for discontinuities in the 116 galaxy clusters of the Cluster HEritage project with XMM-Newton – Mass Assembly and Thermodynamics at the end of structure formation \citep[CHEX-MATE,][]{chexmate} sample. As the name of the project suggests, the images at our disposal were obtained with the XMM telescope as part of a Multi-Year Heritage and are characterized by a high S/N ratio that guarantees a homogeneous reconstruction of the spectral temperature (with a relative error of $\sim$15\%) in the [0.8–1.2] $R_{500}$ annulus. XMM observations are characterized by a lower resolution than those obtained with Chandra, with a half energy width of 13 arcsec and $\sim$0.5--1 arcsec, respectively. However, the data at our disposal, allow us to cover the 116 CHEX-MATE galaxy clusters with a minimum uniform exposure, providing us with the unprecedented opportunity to characterize, from a statistical point of view, the occurrence and the properties of the ICM discontinuities associated with shocks and cold fronts. In this work, we extend the morphological analysis presented in \cite{Campitiello2022} to structures observed in 2D maps, thereby enabling us to identify additional possible proxies for the dynamical state of the CHEX-MATE clusters. Our goal is to provide a preliminary statistical interpretation of the results in the context of the physical processes governing the formation of these structures, including minor and major mergers, sloshing, and AGN outflows. We focus on a systematic characterization of the detected features and their statistical properties, rather than on detailed physical modeling of each system.

The paper is structured as follows: in Sect. \ref{sect:dataset}, we present the sample and the dataset that we analyzed. In Sect. \ref{Sect:method_edges}, we describe the method adopted for the detection of the discontinuities, and we present a comparison with the literature. In Sect. \ref{Sect:catalog_edges}, we present the catalog of the edges detected. In Sect. \ref{Sect:statistical_edge}, we analyze the properties of these edges and their connection with the radio emission. In Sect. \ref{sec:discussion}, we discuss our results and in Sect. \ref{Sect:conclusion}, we draw our conclusions. Throughout the paper, we assume a flat $\Lambda$CDM cosmology with $\Omega_{\rm m} = 0.3$, $\Omega_{\Lambda} = 0.7$ and $H_0 = 70$ km s$^{-1}$ Mpc$^{-1}$. Furthermore, the names of the CHEX-MATE galaxy clusters will be presented without the "PSZ2" prefix coming from the second SZ catalog of sources detected by Planck \citep{Planck205}, for simplicity.

\section{The dataset}\label{sect:dataset}
The CHEX-MATE program \citep[see][for more details]{chexmate} is based on a sample of 118 galaxy clusters detected by Planck through their SZ signal. The sample is divided into two subsamples: Tier 1, consisting of 61 objects located at low redshift in the northern sky (\(0.05 < z < 0.2\) and \(\text{Dec} > 0\)), with masses in the range \(2 \times 10^{14} M_{\odot} < M_{500} < 9 \times 10^{14} M_{\odot}\), providing an unbiased view of the cluster population at the most recent cosmic times; Tier 2, including the most massive systems that have formed thus far in the history of the Universe (\(z < 0.6\) with \(M_{500} > 7.25 \times 10^{14} M_{\odot}\)).
These two subsamples share four clusters. The XMM exposures were tailored to estimate the spectral temperature in the annulus [0.8-1.2] $R_{500}$ with a relative temperature uncertainty of $\sim$15\% statistical uncertainty at $1 \sigma$.
Images were produced using the pipeline developed during the X-COP project \citep{xcop,Ghirardini2019} and adopted by the CHEX-MATE collaboration. In particular, the XMM data were processed using the SAS software (version 16.1.0) and the extended source analysis software (ESAS) package \citep{Snowden2008}. Count-images, exposure maps, and particle background maps are extracted in the narrow [0.7-1.2] keV band, where the ratio between the source and background emission is maximized and, consequently, the systematics related to the subtraction of the EPIC background are minimized \citep{Ettori2010}. A detailed description of the procedure adopted is presented in \cite{Bartalucci23} and a complete gallery of the images is shown in Fig.~6 in \cite{chexmate}. In the CHEX-MATE pipeline, point sources in the observations are identified using the SAS tool \texttt{ewavelet} in two bands ($0.5-2$ keV and $2-7$ keV), and filtered in the $LogN-LogS$ distribution as described in \cite{Ghirardini2019}, to ensure a uniform level of the Cosmic X-ray Background (CXB) emission across the field of view. We further inspected images by eye to identify residual point sources, which could affect our measurements. We masked identified point sources and filled the ``holes'' using an interpolation with the surrounding pixels. We point out that only point sources were masked, so substructures associated with major or minor mergers remain in the images. The image pixel size corresponds to  2.5 arcsec. The final sample is composed of 116 clusters; two (G028.63+50.15 and G283.91+73.87) were excluded due to observational issues \citep[see][for more details]{Bartalucci23}.

\section{The method to detect edges in the X-ray surface brightness map}\label{Sect:method_edges}

The method adopted in this analysis proceeds in two main steps. First, candidate regions potentially hosting SB edges are identified objectively, without assuming a specific merger geometry or requiring supporting evidence from other wavelengths. This initial identification is guided by quantitative criteria applied uniformly across the sample (see Subsect. \ref{subsect:residual_maps}). In the second step, SB profiles are extracted to characterize the features (see Subsect. \ref{subsect:analysis}). This process necessarily involves some manual intervention, as is commonly done in the literature. In particular, the center of the extraction sector, its position angle, and opening angle are manually adjusted to best trace the candidate discontinuity. We choose not to fix the center (e.g., at the X-ray peak) or apply a fully standardized approach, as such choices can lead to systematic misalignment between the sector and the actual feature. This is especially important in clusters with asymmetric morphology or off-center features, where an automated alignment could either miss the edge or yield biased estimates of its properties. The manual refinement thus ensures a more accurate representation of the discontinuity. The analyst-dependent variability introduced at this stage can be mitigated by testing different configurations of the extraction region (see Subsect. \ref{subsect:analysis}, for more details). Moreover, as we will present in detail in the following section, the initial step aims to detect candidate regions through a combination of Gaussian Gradient Magnitude (GGM) filtering and residual analysis. This approach further reduces subjectivity in selecting the curvature, center, and opening angle of the sector compared with a purely visual inspection of the observations. The details of the method adopted are reported in the following Subsections. 

\subsection{Residual maps}\label{subsect:residual_maps}

To identify edges in the ICM distribution, we analyzed the residuals obtained by subtracting the model image representing the expected SB distribution of the considered cluster from the observations. This part of the analysis has been performed by means of the tools provided by two python packages: \texttt{PYPROFFIT}, a package used for the extraction and analysis of SB-profiles \citep[see][for more details]{Eckert2020} and \texttt{GGM} \citep{Sanders2016}, a Gaussian gradient magnitude filter that highlights SB gradients in an image (similarly to the Sobel filter \citep{Sobel1973}, but assuming Gaussian derivatives). The procedure adopted is the following:
\begin{enumerate}
    \item First, we filled the masked point-source regions using the \texttt{dmfilth} function of the \texttt{pyproffit} package. This method is the python version of the \texttt{dmfilth} tool provided by the CIAO\footnote{\url{https://cxc.cfa.harvard.edu/ciao/}} software, and computes a 2D spline interpolation in between the masked regions and generates a Poisson realization of the spline interpolated data, such that the filled holes have similar statistical properties to their surroundings;
    
     \item Starting from this corrected image, we then considered a region of radius $R_{500}$, and we extracted a profile in elliptical annuli centered on the image centroid, with an ellipse axis ratio (major/minor) and position angle calculated with principal component analysis (PCA). In particular, we estimate the centroid and global shape of the X-ray surface brightness by diagonalizing the weighted second-moment matrix of the background-subtracted image within a circular aperture. Each pixel at ($x_i$,$y_i$)
    is assigned a weight $w_i$ proportional to its background-subtracted counts and inversely proportional to the exposure. From the total weights, we compute the weighted centroid ($\bar x$,$\bar y$) and the weighted covariance matrix $\Sigma$. PCA of 
    $\Sigma$ yields eigenvalues $\lambda_1\ge\lambda_2$ and eigenvectors defining the major/minor axes and the position angle. This PCA formulation is equivalent to the classical moment-of-inertia approach widely used for X-ray morphology, with PCA providing a compact linear-algebra view. To account for the complex X-ray morphology that CHEX-MATE clusters may exhibit, we fit the profiles with an elliptical double-$\beta$ model and used the fit results to obtain a model image representing the intrinsic cluster emission. Starting from this model image, we created two types of residual maps:
    \begin{itemize}
        \item The first map, called $R$-map (Fig. A1, \href{https://doi.org/10.5281/zenodo.22715797}{Appendix A} of the attached file, top left panel), is a residual map obtained by subtracting the model image from the original observation of the cluster;
        \item The second map, $GR$-map (Fig. A1, \href{https://doi.org/10.5281/zenodo.22715797}{Appendix A} of the attached file), is built by applying the GGM filter to both the observation and the smooth model, then subtracting the two results. The GGM filter measures the (unsigned) strength of local surface-brightness gradients after Gaussian smoothing at a chosen scale. For this reason, the $GR$-map highlights where the observed gradients are sharper or flatter than expected from the model. By construction, positive GR values mark locations with steeper-than-model edges (candidate edges), negative values indicate flatter-than-model gradients, and values near zero are consistent with the model.
    \end{itemize}
    
    \item Since the distribution of the residuals is approximately Gaussian, we decided to define significant residuals, all those regions showing values higher (or lower) than 3 $\sigma$ (or -3 $\sigma$), where $\sigma$ is the rms of the image. Using this definition, we identified the significant positive (or negative) residuals in both the $R$- and $GR$-maps. The combined use of these two residual maps, along with the adoption of a significance threshold, mitigates the effects of contamination from spurious features, such as those observed to be generated by the GGM due to the low quality of the XMM images (see Sect. \ref{sec:discussion}, for more details).
\end{enumerate}

We represent in Fig. A1 (\href{https://doi.org/10.5281/zenodo.22715797}{Appendix A} in the attached file) the main steps of the procedure adopted and here described.

\begin{figure}
    \centering
    \includegraphics[scale=0.5]{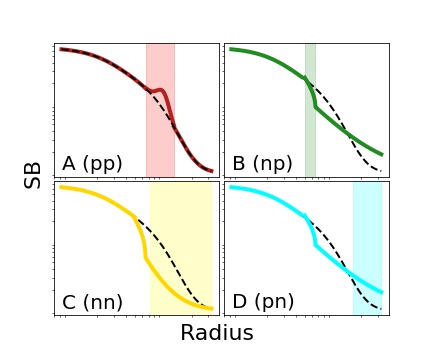}
    \caption{Representation of the four cases presented in Subsect. \ref{subsect:residual_maps}. The dashed line represents the SB distribution model, while the continuous colored line represents the SB expected to be observed in the four cases. In particular, the shaded area in the top left panel shows scenario A ($pp$-regions), in the top right panel shows scenario B ($np$-regions), in the bottom left panel shows scenario C ($nn$-regions), and in the bottom right panel shows scenario D ($pn$-regions).}
    \label{fig:4cases}
\end{figure}

Next, we focused on regions where significant residuals are detected in both the $R-$ and $ GR-$ maps. In particular, we classified these regions in four categories: (A) regions where both the $R$-map residuals and the $GR$-map residuals are positive ($pp$), (B) regions where the $R$-map residuals are negative and the $GR$-map residuals are positive ($np$), (C) regions where the $R$-map residuals are positive and the $GR$-map residuals are negative ($pn$), (D) regions where both the $R$-map residuals and the $GR$-map residuals are negative ($nn$). The four categories of regions identified could be linked to specific features of the ICM, as represented in Fig.~\ref{fig:4cases}.

The $pp$ regions (GR>0 and R>0) identify locations where the observed surface-brightness gradient is sharper than the smooth model and the local surface brightness is in excess. This combination is consistent with the leading boundaries of substructures (e.g., remnant cores, case A in Fig. \ref{fig:4cases}), so we flag $pp$ regions as candidate substructure boundaries, pending confirmation by SB-profile fits (and, when available, thermodynamic diagnostics).

The $np$-regions instead are associated with a steepening of the profile (positive $GR$-map residuals) and a deficiency in the SB distribution (negative $R$-map residuals). These regions are thus good candidates for the detection of edges related to the presence of shocks and cold fronts (case B in Fig. \ref{fig:4cases}), but could also trace the boundaries of cavities. Concerning $nn$-regions, they trace a deficiency in the SB distribution (negative $R$-map residuals) and a flatter profile  (negative $GR$-map residuals), which may highlight the inner regions of cavities (case C in Fig. \ref{fig:4cases}). Finally, it is not clear what the $pn$-regions (case D in Fig. \ref{fig:4cases}) can be attributed to. In our analysis, we will focus not only on the $np$- regions but also on the $pp$- ones. The latter can trace the presence of substructures and may coincide with cold fronts associated with mergers, as these fronts mark the interface between two plasmas with different temperatures. At the same time, even in the case of sloshing, we expect the ICM to exhibit both excesses and deficits in surface brightness, as these features trace the characteristic pattern of the phenomenon.

\subsection{Analysis of the residuals}\label{subsect:analysis}

While ICM edges can be more readily detected in the higher angular resolution Chandra images, XMM provides dramatically larger effective area and larger FOV and hence enables better characterization and better survey of cluster structures. We use the residual-based pre-selection, described above, to quantitatively characterize structures and features. In practice, the $R$ and $GR$ maps act as data-driven proxies for visual inspection, highlighting the areas where gradients are sharper than the smooth model and/or the SB is in excess, and thus indicating where to concentrate the subsequent, profile-based edge search. 

From this point onward, the analysis follows a standard workflow: we define arc-aligned extraction sectors to follow either the arc-like morphology traced by the candidate regions highlighted in the $R$/$GR$ maps ($pp$ or $np$, as defined above) or their boundaries, and then extract the corresponding surface-brightness profiles. We used the same approach reported in \cite{Botteon2016b, Botteon2018}: the residual maps provide a good starting point for delineating the sector. We then adopted different apertures, radial ranges, and positions, and selected those that maximize the jump, as indicated by the best-fitting statistics. Although this approach necessarily involves subjective decisions, we mitigate this by exploring a range of sector configurations. The detailed sector parameters, including center positions and angles used for all detected edges, are provided in the Table B1 (of the \href{https://doi.org/10.5281/zenodo.22715797}{Appendix B} in the attached file) to ensure reproducibility.

The SB profiles of the candidate shocks and cold fronts were then modeled under the assumption that the underlying density profile follows a broken power law \citep[e.g.,][and references therein]{Markevitch2007}. In the case of spherical symmetry, the inner and outer (subscripts $i$ and $o$) densities differ by a compression factor $\mathcal{J}=n_i/n_o$ at the location of the jump $d_j$:

\begin{equation}
\begin{aligned}
n_i(d) &= \mathcal{J} n_0 \bigg( \frac{d}{d_j}\bigg)^{\alpha_1}  \, \, \, \text{if} \,\,\,  d\leq d_j  \\
n_o(d) &= n_0 \bigg( \frac{d}{d_j}\bigg)^{\alpha_2}  \, \, \, \, \,\,\, \text{if} \,\,\,  d> d_j
\end{aligned}
\end{equation}
where $\alpha_1$ and $\alpha_2$ are the power-law indices, $n_0$ is a normalization factor and $d$ denotes the distance from the center of the sector. All these quantities were free to vary during the fitting procedure. As a final step, we inspected the results of the fits by eye, to check for the shape of the profile identified. In particular, we expect to find a shape similar to the one presented in Fig. \ref{fig:4cases}, panel B (green line). Edges with negative slopes in the inner region, or with a very smooth profile with no indication of a jump and $\mathcal{J}\leq1.5$ from the fit, will be listed in our catalog as "possible discontinuities" and will not be considered in the statistical analysis. We emphasize that $\mathcal{J}\geq 1.5$ is not used as a hard selection criterion. Rather, it is considered only for candidate edges whose surface-brightness profiles do not show the presence of a jump. Well defined edges with $\mathcal{J} \leq 1.5$ are instead included in the final catalog. We note that real clusters frequently display ellipsoidal or complex morphologies that may bias the observed profiles and derived parameters due to projection effects. However, \texttt{pyproffit} mitigates this issue through a deprojection procedure that provides a three-dimensional estimate of the density jump.

\subsection{Comparison with the literature}\label{Subsect:comparison_Chandra}
\begin{figure*}
    \centering
    \includegraphics[scale=0.6]{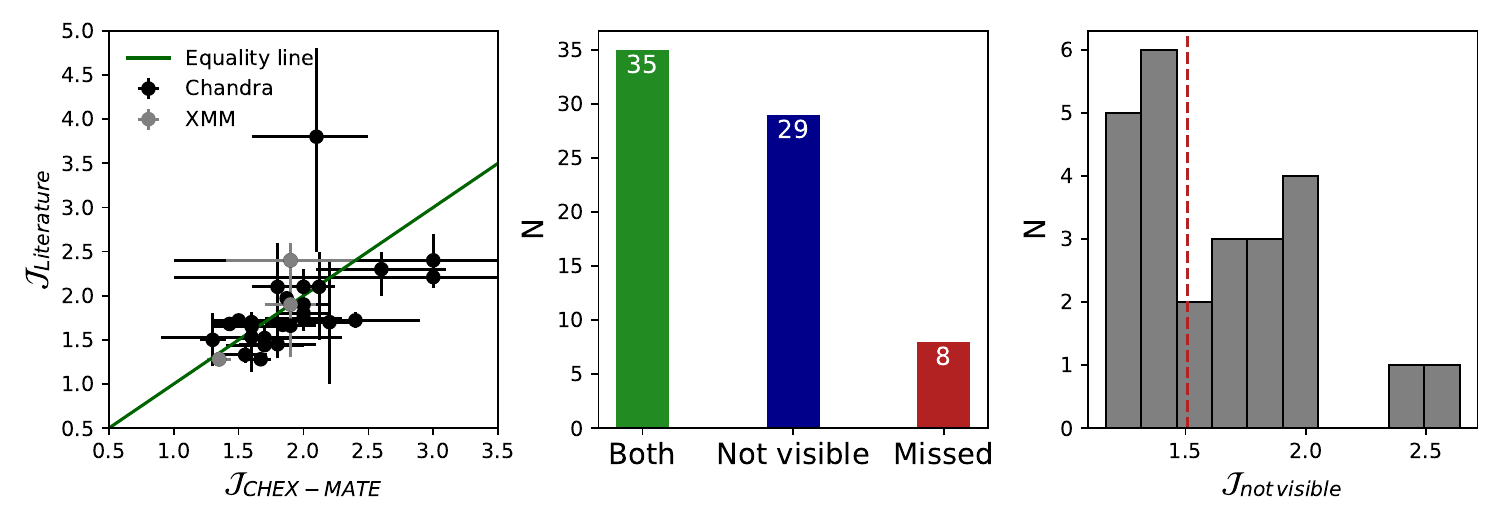}
    \caption{Left: Comparison of jump values ($\mathcal{J}$) from the literature, measured with XMM (gray points) and Chandra (black points), and those obtained in this work. Center: Distribution of literature edges detected by both Chandra and our method (`Both'), edges not visible in the XMM surface-brightness profiles (`Not visible'), and edges visible in XMM but missed by our method (`Missed'). Right: $\mathcal{J}$ distribution of the literature edges not recovered in our analysis. Only edges with published $\mathcal{J}$ measurements are shown. In these cases, the XMM surface-brightness profiles extracted at the reported edge locations do not show a significant discontinuity.}
    \label{fig:Chandra_comparison}
\end{figure*}

In this section, we conducted a comprehensive comparison of our results with all edges of the CHEX-MATE clusters previously detected and analyzed in the literature, to assess the robustness of our method, and the reliability of our detection. In the attached file, \href{https://doi.org/10.5281/zenodo.22715797}{Appendix C} (see Tab. C1), we provide a detailed summary of this check, including citations to prior studies and potential reasons for missed detections. Where available, we compared the magnitude of the jump, $\mathcal{J}$, measured in our analysis with estimates from previous studies. The majority of these detections and estimates were derived from Chandra observations, which, as highlighted in the introduction, offer superior angular resolution, ideal for such analyses, making it particularly well-suited for detecting sharp surface brightness edges. Chandra has a PSF with a half-power diameter (HPD) of approximately 0.5 arcseconds on-axis, whereas XMM's PSF has an HPD of about 15 arcseconds. This difference in resolution affects the ability to resolve narrow features, with XMM being more prone to smoothing out fine structures.
The results of this comparison are illustrated in Fig.~\ref{fig:Chandra_comparison} (left panel), where we show the magnitude of the jumps reported in the literature (y-axis) with the magnitude measured in our work (x-axis). It is evident that for most edges, the magnitude of the jump measured by XMM is consistent with those measured by Chandra. The central panel of Fig. \ref{fig:Chandra_comparison} illustrates the number of successful and missed detections. We found that our method detected 35 out of 72 ($\sim$49\%) of edges documented in the literature. For the remaining approximately 51 \% of missed detections, we examined profiles across the expected discontinuity regions. In 29 out of 72 cases ($\sim$40\%), no discontinuity was visible in the XMM profiles. The median density jump of these edges that are not visible in the XMM images is $\mathcal{J}=1.51$. Figure~\ref{fig:Chandra_comparison} (right panel) shows the distribution of $\mathcal{J}$ for this subsample. 
We further investigated whether the missed detections are randomly distributed across the sample, by examining how the recovery fraction varies with the dynamical state, and between the Tier 1 and Tier 2 objects. For each cluster with at least one edge reported in the literature, we computed the ratio between missed detection and the total number of known edges. We then considered the average of this ration for each cluster category. We found that relaxed systems are characterized by a significantly lower recovery fraction than mixed and disturbed systems: only $\sim$25\% of the edges in relaxed systems are recovered, compared to the 63\% and 70\% of the mixed and disturbed classes, respectively. By contrast, no significant difference is observed between Tier 1 and Tier 2 clusters, for which the average recovery fractions are $\sim56\%$ and $\sim52\%$, respectively. To better understand the origin of this bias in the relaxed systems, we inspected the properties of the missed detections reported in the literature. Nearly all missed edges (9/10) are either located close to the cluster core, or characterized by low $\mathcal{J}$ values (i.e., lower than 1.5). This suggests that the dominant source of incompleteness for this class of clusters is the loss of weak central cold fronts, likely due to PSF smearing and limited angular resolution. 

In only 8 cases ($\sim$11\%), the edge was visible in the profile but not detected by our procedure (red column in Fig. \ref{fig:Chandra_comparison}, central panel). Among the clusters with edges known in literature, there are three objects that were observed with XMM. The first case is the shock in G008.94-81.22 \citep[Abell 2744][]{Eckert2016}, for which we retrieve the same jump value reported in the paper. The second cluster, G044.20+48.66 \citep[Abell 2142,][]{Rossetti2013}, exhibits four cold fronts. Among these, we detect only the one on the western side, which appears sharper in the images and is closer to the cluster center. We missed the detection of the inner and southern cold fronts shown in \cite{Wang2018}, probably due to PSF effects and we do not detect the southeastern cold front. This may be because it is located in the cluster outskirts, where the signal-to-noise ratio is lower and the surface brightness contrast may be weaker. Since our approach primarily identifies regions with the most significant residuals (exceeding $\pm 3\sigma$), it is possible that we miss the position of this second edge, whereas a more detailed analysis focused specifically on this cluster has detected it. The last case is G208.80-30.67 \citep[Abell 521][]{Bourdin2013}, a highly disturbed system. None of the four features identified in the previous studies was detected by our pipeline. By extracting profiles across the regions where these features were reported in Abell 521, we do not find an edge corresponding to the eastern and western shocks or the northern cold front. However, we detect a discontinuity at the southern cold front, which we classify as a missed detection. The discrepancy between these two analyses may be due to different treatments of the background and different exposure times for the X-ray maps used.
We note that detecting edges in XMM images is intrinsically challenging. The method presented in this paper aims to overcome this limitation. We use a residual-based pre-selection ($R$ and $GR$ maps) to highlight the most promising edge locations in a uniform, data-driven way across wide fields, and then extract SB profiles only in those candidate regions. The comparison with the literature indicates that: (i) most non-detections are attributable to the limitations of the XMM data rather than to our workflow; and (ii) restricting the search to locally significant residuals sacrifices only a small fraction of edges that would require object-by-object, highly tailored analyses. We considered this an acceptable trade-off for a large, homogeneous sample. We do not aim to discover new features in well-studied objects (many of which were characterized with deep Chandra data). Instead, our goal is a conservative census: the method recovers the bulk of known edges over an extended sample with XMM, while providing a consistent signed taxonomy ($pp/np/pn/nn$) and a standardized path to subsequent, object-specific follow-up where warranted.The exquisite spatial resolution of Chandra, while ideal for detailed studies of individual clusters, makes large-population analyses extremely resource-intensive. This is especially true now that Chandra’s effective area has dropped significantly compared to the beginning of the mission, further limiting its efficiency for systematic surveys. No future missions with Chandra-like resolution are currently planned. Thus, exploiting XMM’s slightly lower resolution is presently the best strategy for systematic studies over extended samples. The results of this comparison provide an overview of the limitations of our analysis that will be further discussed in Sect.~\ref{sec:discussion}.
In addition to this comparison, we investigated the role that image noise may play in edge identification. This analysis and its results are shown in the \href{https://doi.org/10.5281/zenodo.22715797}{Appendix D} of the attached file.

\begin{figure}
    \centering
    \includegraphics[scale=0.64]{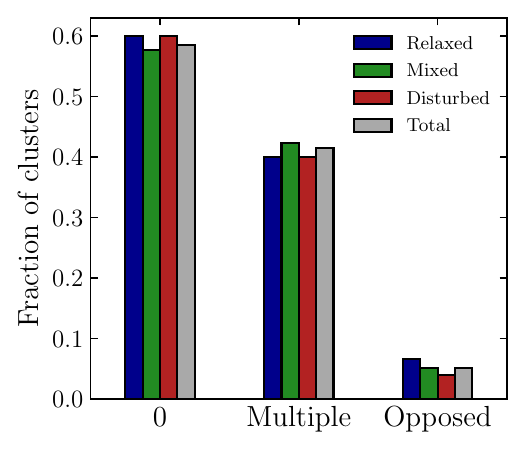}
    \caption{Histogram of the fraction of clusters with one or multiple edges (“Multiple”) or no edges (“0”), shown by dynamical state—relaxed (blue), mixed (green), disturbed (red)—and for the full sample (gray). The rightmost column reports the fraction of “Opposed” edges. The incidence of edges is broadly similar across dynamical states.}
    \label{fig:catalog}
\end{figure}

\begin{figure}
    \centering
    \includegraphics[scale=0.9]{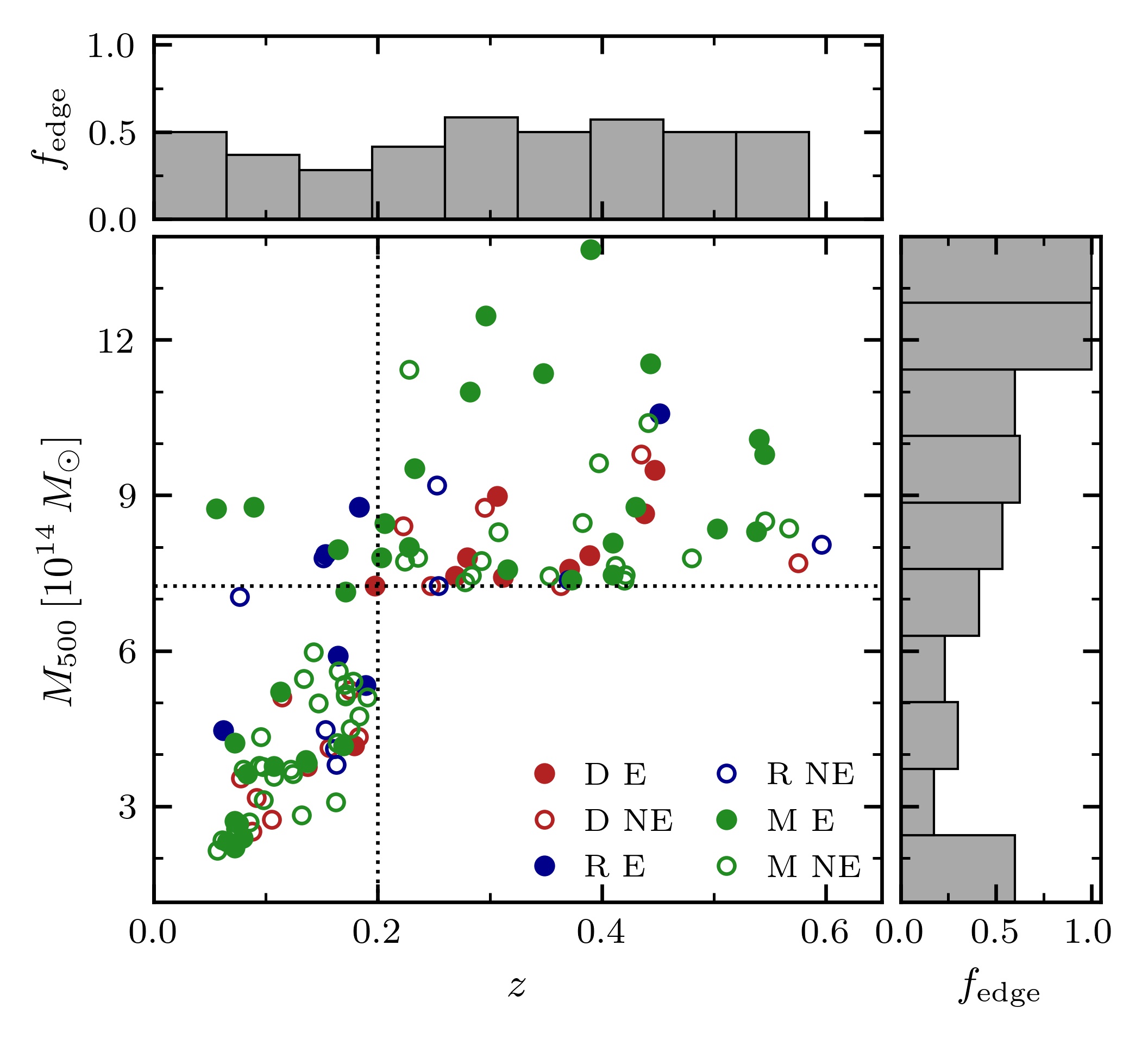}
    \caption{Distribution of edges in the \(M_{500}-z\) CHEX-MATE plane. As described in Sect.\ref{sect:dataset}, CHEX-MATE groups together a subsample of the most recently formed clusters in the Universe (Tier 1) and a subsample of the most massive clusters (Tier 2). Filled markers indicate clusters with at least one edge (E), while empty markers represent clusters without edges (NE). The markers are color-coded by cluster dynamical state: relaxed (R, blue), mixed (M, green), and disturbed (D, red). The vertical and horizontal lines delineate the Tier 1 and Tier 2 samples. The top and right panels show the fraction of clusters with at least one detected X-ray edge as a function of z and M$_{500}$, respectively, computed as the ratio between the number of clusters with an edge and the total number of clusters in each bin. }
    \label{fig:edge_dist}
\end{figure}

\section{The catalog of edges in SB maps}\label{Sect:catalog_edges}

Using the procedure above, we identify 66 edges, 32 of them newly detected, in 48 clusters. Edge properties are reported in Table E1 (\href{https://doi.org/10.5281/zenodo.22715797}{Appendix E} in the attached file). In the gallery shown in Fig. F1 (\href{https://doi.org/10.5281/zenodo.22715797}{Appendix F} in the file attached), solid black arcs mark edges that pass the final inspection of the SB profile, while dashed arcs indicate features that fail our criteria and are not retained. We discard the latter because their atypical profiles preclude a secure characterization and could bias the results. The corresponding SB profiles are shown in Fig. G1 of the \href{https://doi.org/10.5281/zenodo.22715797}{Appendix G} in the attached file.

To assess whether adjacent arc-like features belong to the same edge or to distinct edges, we apply the following consistency checks. First, we extract the profile in an arc-aligned sector wide enough to encompass both candidates and verify whether the discontinuity is still identified by the fit; if so, we treat them as one edge. If a single-sector fit is not adequate (e.g., because the curvature varies), we extract profiles in a mosaic of adjacent, narrower sectors that follow the local curvature and compare the best fit values of the discontinuity radial distance, $d_x$, and compression, $\mathcal{J}$, obtained for the various sectors. Two segments are considered the same edge if both their $d_x$ and $\mathcal{J}$ values agree within their 1$\sigma$ uncertainties. In that case, we report the segments separately in Table B1 of the attached file but tag them with a superscript '$^\circ$' to indicate that they are treated as a single edge in the subsequent analysis; the radius and jump used downstream are the inverse-variance–weighted means of the individual measurements. If these conditions are not met, we classify the segments as different edges. As illustrative cases, in G229.74+77.96 the three eastern segments satisfy our consistency checks and are treated as a single discontinuity. By contrast, in G049.22+30.87 three arcs delineate the characteristic spiral pattern of gas sloshing and fail the consistency criteria in both $d_x$ and compression $\mathcal{J}$; we therefore regard them as distinct edges.  

In a few clusters (G073.97-27.82, G092.71+73.46, G159.91-73.50, G172.98-53.55, G325.70+17.34), sectors drawn across an $np$ patch yield a best-fit discontinuity at a slightly larger radius, i.e. just beyond the $np$ region. In all such cases, the $np$ patch is immediately followed azimuthally by an $nn$ patch (surface-brightness deficit with a flat gradient). We retain these features in our sample because their arcs are spatially coherent with the $np$ regions and because, as noted above, while residuals guide the sector definition, the fit ultimately confirms the detection. We flag these identifications in the catalog as we acknowledge that they are less robust than typical cases. In any case, we note that their inclusion does not affect the statistical conclusions of this work. Regarding peculiar features, in a few systems we identified $np$-regions (often accompanied by $pp$-regions) that appear to trace a spiral shape. Examples include G057.78+52.32, G077.90-26.63, G098.44+56.59, G186.37+37.26 and G243.15-73.84. This pattern may indicate sloshing and warrant further analysis with higher-angular-resolution Chandra data. 

In Fig.~\ref{fig:catalog}, we summarize our results: 
48 CHEX-MATE clusters ($\approx$41\%) show one or more edges.
The incidence of edges does not vary significantly with dynamical state. The fractions of systems with zero or multiple (i.e., >1) edges are nearly the same for relaxed, disturbed, and mixed clusters \citep[as defined by][]{Campitiello2022}. Likewise, the occurrence of opposed edges (i.e. edges located on opposite sides of the core) shows no clear trend with dynamical state. In Fig. \ref{fig:edge_dist}, we present the distribution of clusters with and without edges in the $M_{500}$–$z$ plane. It can be observed that the high- and low-mass regimes, as well as the high- and low-redshift regimes, are populated differently. Specifically, we find that the fraction of edges detected is higher in systems with masses greater than $7 \cdot 10^{14} M_{\odot}$ ($\sim 54\%$) compared to those with masses below this threshold ($\sim 27\%$). Similarly, clusters at $z > 0.2$ show a higher fraction of edges ($\sim 50\%$) than those at lower redshifts ($\sim 34\%$). However, mass and redshift are strongly coupled in our sample by construction, since the most massive clusters predominantly lie at higher redshifts. As a consequence, the observed trends cannot be disentangled and may reflect the same underlying effect. In general, this trend shows that the Tier 2 subsample has more edges than the Tier 1 subsample, despite the fact that spatial resolution worsens at higher redshifts. All the features presented in this paper are measured within $R_{500}$; outer features would not be detectable with our data. We therefore avoid drawing global conclusions about the overall merger rate.

\section{Statistical properties of the edges} \label{Sect:statistical_edge}

\begin{figure}
\centering
\includegraphics[scale=0.89]{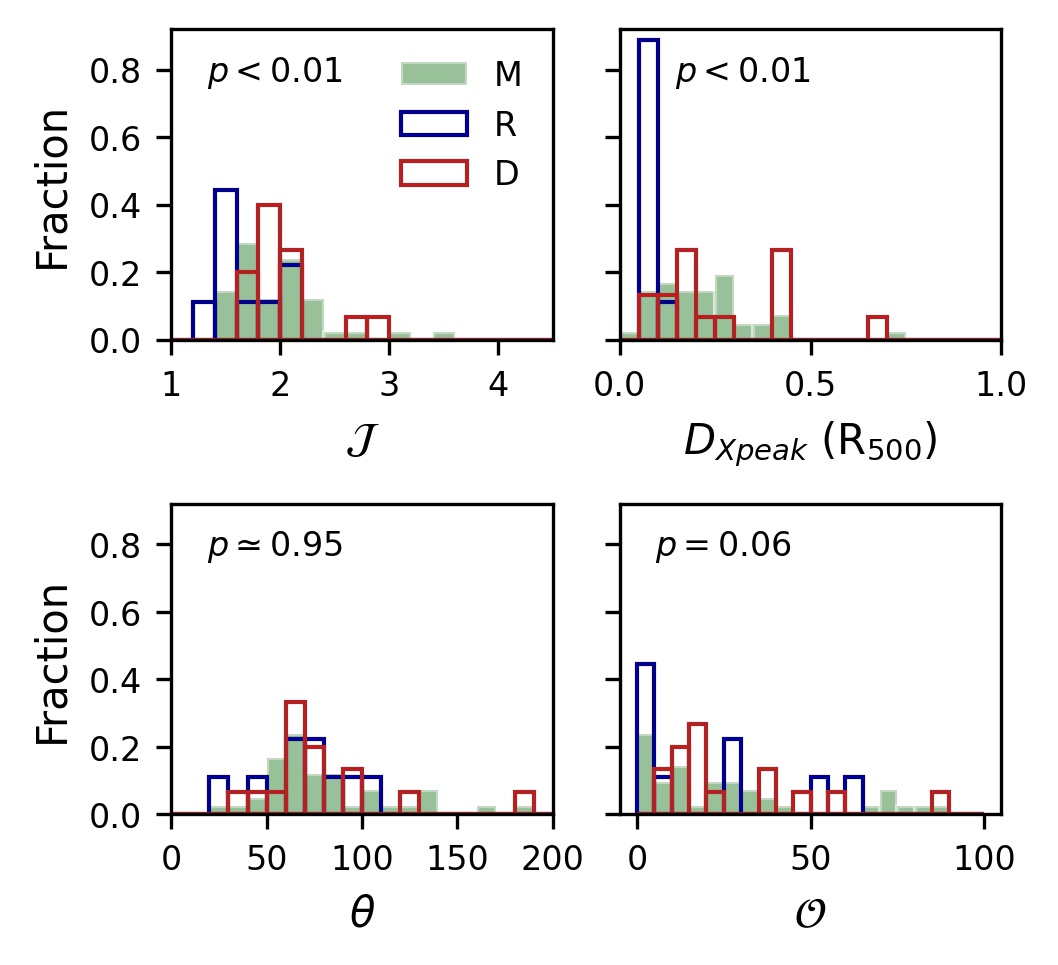}
\caption{Distribution of the jump ($J$), the distance from the X-ray peak ($D_{Xpeak}$) in units of $R_{500}$, the angular extension ($\theta$), and the orientation of the edge relative to the ellipse that best approximates the X-ray morphology of the cluster ($\mathcal{O}$). For the definition of these parameters, refer to Sect. \ref{Sect:statistical_edge}. Colors indicate the dynamical state: blue for relaxed (R), red for disturbed (D), and green for mixed (M) systems. }
\label{fig:cumulative}
\end{figure}

For each edge detected, we measured the magnitude of the jump (as described in Subsection \ref{subsect:analysis}), $\mathcal{J}$, its distance from the X-ray peak, $d_X$, its angular extension, $\theta$, and its orientation $\mathcal{O}$. 
For this latter parameter, we proceed as follows. We first determine the edge position angles at the arc's center, defined as the mid-angle of its azimuthal extent. Angles are meadured in degrees from the R.A. (x) axis, increasing counter-clockwise. We then measure the cluster reference angle as the position angle of the best-fitting ellipse, i.e. the orientation of its major axis in degress from the R.A. axis. Due to its definition, $\mathcal{O}$ was computed only for those clusters whose X-ray morphology has the shape of an ellipse with axial ratio higher than 1.1. The distributions of all these quantities are shown in Fig. \ref{fig:cumulative}. The vertical lines show the median values obtained for the relaxed (blue), mixed (green), and disturbed (red) populations.

Among these parameters, the angular size is the only one that does not show significant differences between relaxed and disturbed systems. In contrast, the median values of the other parameters - $\mathcal{J}$, $d_X$, and $\mathcal{O}$ - differ between these populations. Specifically, we find $\mathcal{J}_R = 1.5$, $\mathcal{J}_D = 1.9$ for relaxed and disturbed objects, respectively; $d_{X,R} = 0.073$ and $d_{X,D} = 0.19$ (in units of $R_{500}$); and $\mathcal{O}_R = 26$, $\mathcal{O}_D = 16$ degrees. To assess differences between relaxed and disturbed clusters, we applied a two-sample Kolmogorov–Smirnov (KS) test.  For the compression factor, 
$\mathcal{J}$, and distance from the X-ray peak, $d_x$, we obtain $p<0.01$ indicating statistically significant differences between the distributions. For the orientation parameter $\mathcal{O}$, $p=0.06$, just above the conventional significance threshold ($p<0.05$) suggesting a possible difference that warrants further investigation. For the azimuthal extent $\theta$, $p\simeq 0.95$, consistent with no difference between the two populations. These results suggest that edges in relaxed and disturbed clusters likely originate from distinct processes, with edges in relaxed systems tending to be closer to the X-ray peak, shallower, and less aligned with the overall X-ray morphology of the clusters, as expected in the case of sloshing.

\begin{figure}
\centering
    \includegraphics[scale=0.69]{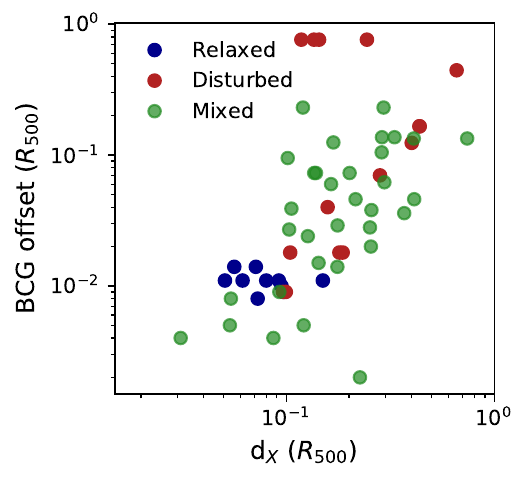}
    \caption{Correlation between the BCG-offset and the distance of the edge from the X-ray peak (in units of $R_{500}$).}
    \label{fig:enter-label}
\end{figure}

\begin{figure}
   \centering
    \includegraphics[scale=0.89]{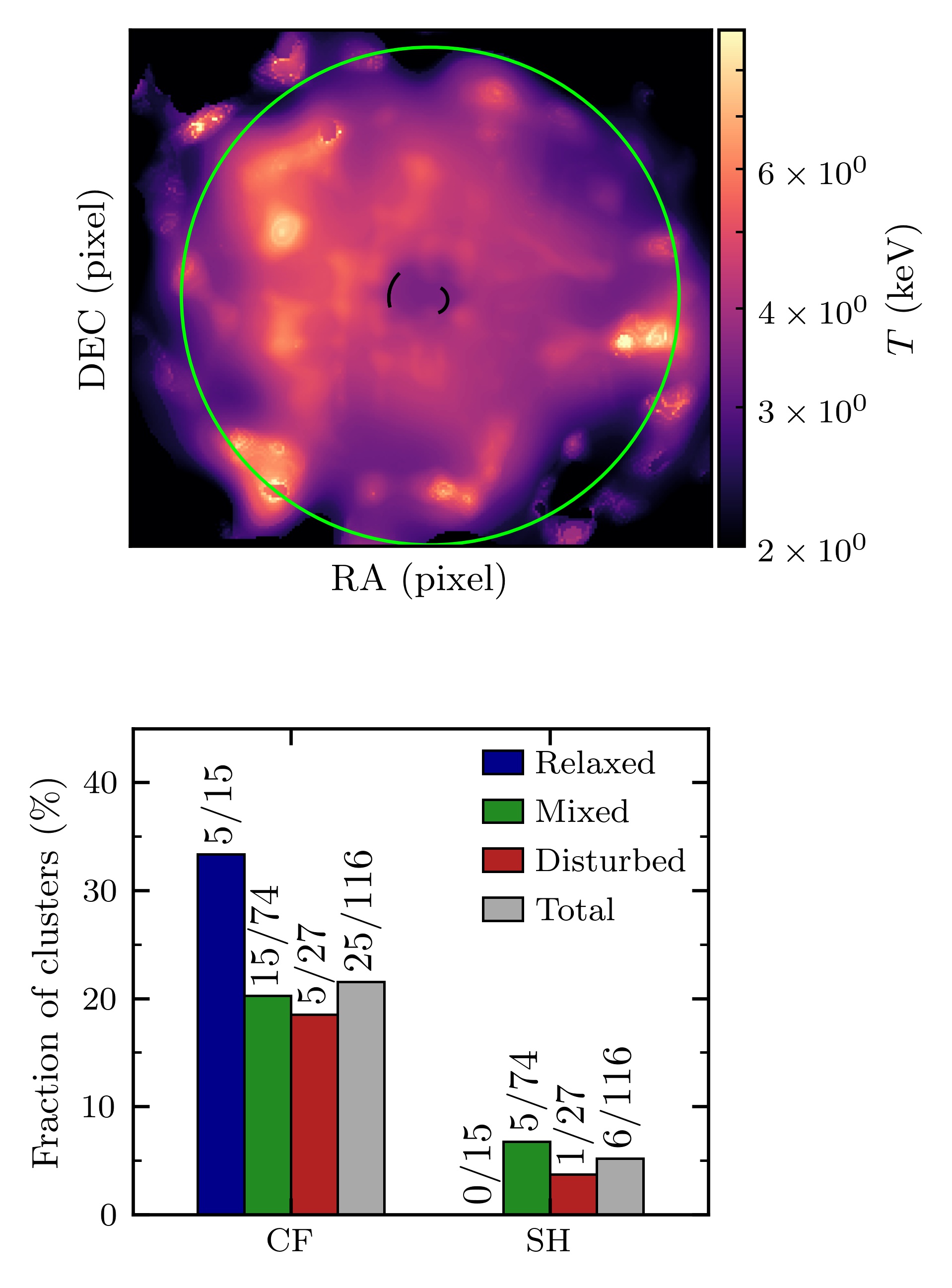}
    \caption{Top: The curvelet temperature map of the cluster G057.92+27.64 is reported here as an example. The green circle has a radius equal to $0.5 R_{500}$. The black arches represent the position of the edges detected by our method. These edges were already known in the literature \citep{Ettori2013}. Bottom: Fraction of clusters hosting at least one cold front (CF) or shock (S) in each dynamical class and in the full sample. Labels report the number of hosting clusters relative to the total number of clusters in each category.}
    \label{fig:Tmap}
\end{figure}

\subsection{Correlation with   X-ray morphological parameters}\label{Subsect:morpho_param_edges}

We examined whether clusters with and without edges exhibited different distributions of the morphological parameters. In particular, we considered the concentration $c$, the centroid shift 
$w$, the power ratios $P_{20}$ and $P_{30}$, and the ellipticity, $e$. These quantities are commonly used to quantify deviations from the regular and approximately symmetric X-ray morphology expected in relaxed clusters. $c$ is defined as the ratio between the surface brightness measured within two concentric apertures, and describes how peaked is the X-ray emission. $w$ quantifies how the distance between the X-ray centroid and X-ray peak positions vary, when measured within apertures of increasing size, tracing large-scale asymmetries and dynamical disturbances. The power ratios $P_{20}$ and $P_{30}$ are multipole moments of the X-ray surface brightness distribution and are sensitive to ellipticity and substructures, respectively. Finally, $e$ measures the deviation of the X-ray emission from circular symmetry \citep[see][for more details]{Campitiello2022}. Additionally, we computed and included the BCG-offset, which quantifies the displacement between the position of the brightest cluster galaxy (BCG) and the X-ray peak, expressed in units of $R_{500}$. This parameter is expected to serve as an indicator of the disturbance level in both disturbed and relaxed systems. We excluded two clusters whose reported BCG offsets exceeded $R_{500}$ and could not be reliably validated. Since these clusters host a total of three edges, this corresponds to the exclusion of three data points from the correlation analysis. Results are shown in Fig. H1 (\href{https://doi.org/10.5281/zenodo.22715797}{Appendix H} in the attached file). The statistical test used for these comparisons was the KS test. We found that only disturbed clusters with and without edges showed different distributions for $P_{20}$, $P_{30}$, and $e$, with $p$-values being less than 0.01, close to 0.05 and less than 0.05, respectively. All other comparisons showed no significant differences between the two groups. Furthermore, we investigated potential correlations between edge parameters and morphological parameters. A strong correlation was found between the BCG-offset and the distance from the X-peak, with a Spearman correlation coefficient of 0.66. This correlation is shown in Fig.~\ref{fig:enter-label}.

\subsection{Classification of the edges}\label{Subsect:classification}

To determine whether the detected edges correspond to shocks or cold fronts, it is essential to analyze the temperature and pressure trends across the edges. Extracting detailed temperature profiles for all edges identified in XMM observations requires defining extraction regions with sufficient counts and can be computationally expensive when repeated across tens of edges across many clusters. Therefore, we opted to use the curvelet temperature maps produced with a customized version of the spectral-imaging SZ algorithm introduced by \cite{Bourdin2015}. This algorithm operates as follows. First, log-likelihood estimates of the ICM’s projected temperature are combined with a curvelet analysis. These temperature log-likelihoods are spatially weighted at each map pixel using B3-spline wavelet kernels that account for both positive and negative contributions. The wavelet and curvelet transforms then facilitate the extraction of temperature information. Specifically, temperature features are derived from the wavelet coefficients, while spatially weighted Fisher information provides a measure of the expected fluctuations. The final temperature maps are reconstructed from denoised curvelet transforms, applying a 4$\sigma$ threshold to the curvelet coefficients. The detected temperature features typically span scales ranging from 3.5 to 60 arcseconds. The methodology used to derive log-likelihood estimates of the projected ICM temperature was originally introduced in \cite{Bourdin2004} and subsequently applied in, e.g., \cite{Bourdin2008}. The main advantage of this method is that it provides direct temperature maps without assuming any a priori geometry for the underlying structures, in contrast to traditional analyses based on radial binning across surface-brightness edges, which may be sub-optimal for following curved or irregular features. In this context, spatial filtering techniques such as wavelet or curvelet transforms can be used to denoise the temperature maps and enhance structures over a range of spatial scales; examples of curvelet-based applications include the CHEX--MATE analysis of SPT clusters \citep{Oppizzi2023} and the study of A521 by \cite{Santra2024}. As an example, Fig.~\ref{fig:Tmap} (left panel) presents the temperature map generated for the cluster G057.92+27.64. Uncertainties in these maps are estimated by performing 100 bootstrap realizations. Generally, the uncertainties depend on the signal-to-noise (S/N) ratio of the observations. From the X-ray peak to the cluster outskirts, the typical 1$\sigma$ relative uncertainty ranges from 4\% to 50\%. These uncertainties do not permit definitive conclusions about the nature of the edges in the ICM. For this reason,  we emphasize that these maps provide qualitative guidance rather than definitive temperature measurements, and classifications based on them should be considered preliminary. The temperature maps are shown in Fig. I1 of \href{https://doi.org/10.5281/zenodo.22715797}{Appendix I} in the attached file.

Based on this classification, we identified 8 potential shocks and 34 potential cold fronts (hereafter shocks and cold fronts), while the remaining edges could not be classified. The bottom panel of Fig.~\ref{fig:Tmap} shows the fraction of clusters hosting at least one cold front or shock in each dynamical class and in the full sample, while the corresponding numbers of classified features are reported in Table~H.2 of \href{https://doi.org/10.5281/zenodo.22715797}{Appendix~H} in the attached file. Cold fronts occur in comparable fractions in mixed and disturbed clusters ($\sim$20\% and $\sim$19\%), while shocks are rare in both classes ($\sim$7\% and $\sim$4\%, respectively) and absent from relaxed systems. This finding is consistent with the expectation that edges in relaxed systems are typically linked to sloshing phenomena. It confirms that the perturbations observed in the ICM of these systems are minor, suggesting that either the perturber has already moved beyond the field of view or the events occurred long ago, with sloshing persisting over extended timescales.

We performed a KS test to assess whether the distribution of parameters characterizing the edges (i.e., $\mathcal{J}$, $d_x$, $\theta$ and $\mathcal{O}$) differs between shocks and cold fronts. The results are summarized in Table H1 of the attached file (\href{https://doi.org/10.5281/zenodo.22715797}{Appendix H}). We found a significant difference between the distributions of $d_X$ and $\mathcal{O}$, with p-values $p=0.018$ and $p=0.049$, respectively. In particular, shocks tend to be located at larger distances from the X-ray peak and to be less aligned with the global morphology of the X-ray emission. No peculiar trend was instead observed for $\mathcal{J}$ and $\theta$. Given the small shock sample ($N=8$), these results should be regarded as tentative, because the statistical power of the comparison is limited. Additionally, we performed a KS test to compare the morphological parameters of clusters hosting shocks and cold fronts. We did not find significant differences between the two populations. 

\subsection{Presence of edges and connection with the radio emission}\label{Subsect:radio_emission_edges}

We tested whether edges and diffuse radio emission occur in the same systems, without requiring any spatial coincidence between the X-ray discontinuity and the radio feature. Specifically, we focused on the presence of radio halos (RH, this category includes also giant radio halos, gRH), relics (RR), and mini-halos (MH), which, like edges, are expected in clusters with turbulence induced by mergers (radio halos and relics) or sloshing \citep[mini-halos, typically observed in relaxed clusters, e.g.][for a review on the topic]{vanweeren2019}. Of the 118 CHEX-MATE clusters, 103 have radio coverage; 77 of these host diffuse radio emission (RH, MH or RR), and two a not classified diffuse emission (DE).

Among the 48 clusters with at least one edge, 38 also show diffuse radio emission. Five of the remaining 10 lack radio data; restricting to the 43 edge–clusters with radio coverage, $38/43 \approx$ 88\% host radio emission. Viewed from the reverse perspective, among the 79 radio–loud systems, 38 (48\%) exhibit at least one edge. The fraction of radio–loud clusters that host edges is broadly similar across RH/RH+RR classes ($\sim$50–60\% see Table \ref{Tab:radio_CFS}), and across dynamical states ($\sim 45-50 \%$ for relaxed, mixed and disturbed systems).

We did not require radio–X-ray alignment, but we note that we did not identify edges clearly co-spatial with relics as it is typically expected. On the other hand, given that this type of diffuse radio emission lies in the outskirts, this is plausibly due to lower S/N and stronger PSF and vignetting effects at large radii, which hamper shock detection there.
\begin{table}
\caption{Connection between X-ray edges and diffuse radio emission.} \label{Tab:radio_CFS}
\centering
\begin{tabular}{@{}lcccc@{}}
\toprule
 & Clusters with edges / Total & CF & S & NI  \\ \midrule
Radio halo & $23/45$ &  16 & 4 & 10   \\
Relic  & $2/5$ & 2 & - & 1  \\
Relic + Radio halo & $11/19$ & 4 & 3 & 8  \\ 
Mini halo & $2/8$  & 3 & - & 1\\
\midrule
\end{tabular}
\tablefoot{Column 2 gives the number of clusters with at least one X-ray edge over the total number of clusters hosting each radio-source type; the remaining columns list cold fronts (CF), shocks (S), and unclassified edges (NI). The absence of shocks in relic-hosting systems may reflect the limited sensitivity of the XMM data in the cluster outskirts, where relics are typically found. For mini-halo systems, the absence of shocks is consistent with the sloshing scenario commonly associated with these sources.}
\end{table}

\section{Discussion}\label{sec:discussion}

In spirit, our pipeline follows a GGM-style edge–enhancement approach; however, applying GGM or other image-space edge finders, such as the Canny algorithm \citep{Canny1986}, directly to XMM images yields a very noisy field of candidate features due to the broader PSF and lower per-pixel S/N. Instead, we combine the intensity residuals (data minus smooth elliptical model) with the gradient residuals (GGM-filtered data minus GGM-filtered model; Sect. \ref{Sect:method_edges}). This suppresses large-scale trends and reduces noise, providing an objective pre-selection of regions to investigate for edge presence.

Using this strategy, we searched for SB discontinuities in the CHEX-MATE sample. The XMM maps at our disposal allow, for the first time, a statistical census of edge occurrence and properties across a large, SZ-selected sample with homogeneous X-ray coverage out to $R_{500}$ and spanning wide ranges in redshift and mass. Detecting edges with XMM remains challenging given its resolution, and we therefore assessed these limitations by comparing our results to the literature (Sect.~\ref{Subsect:comparison_Chandra}). We estimate that our workflow misses $\sim11\%$ of edges that should be detectable in the XMM data (8/72): these are present in the images but not selected by our residual-based pre-selection. The missed cases occur in highly disturbed systems with complex morphologies, where subtracting a (possibly elliptical) symmetric model can leave structured residuals that either fragment a feature or hide it. 

Separately from methodology, XMM’s observational limits are important: about 40\% of the edges reported with Chandra are not visible in SB profiles extracted from the same regions in the XMM maps, indicating resolution/SB constraints rather than a failure of the pre-selection. Nearly half of these non-recoveries lie near the cores of relaxed/mixed systems and are plausibly smeared by the broader PSF. Taken together, these effects imply that our measured incidence is conservative; the true fraction of CHEX-MATE clusters hosting edges could be substantially higher. The edge-level recovery fractions derived from the literature comparison (Sect. \ref{Subsect:comparison_Chandra}) cannot be translated into a correction of the cluster incidence, because they do not constrain how missed edges are distributed among clusters. A cluster-level estimate is likewise unreliable due to the small literature samples and would yield an unphysical incidence above 100\% for relaxed clusters if naively applied. A robust correction would require dedicated mock observations reproducing the selection effects of our analysis, which are beyond the scope of this work. For this reason, the results presented here should be interpreted as a census of edges detectable in XMM data using a uniform analysis procedure, rather than as a complete inventory of all ICM edges in the CHEX-MATE clusters.

To further assess selection and projection effects, we compare our edge census against the available radio data for CHEX–MATE. Major mergers, core sloshing, and AGN feedback produce X-ray edges (shocks and cold fronts), and also stir the ICM, driving turbulence that can (re-)accelerate relativistic particles and, in the presence of $\mu G$ magnetic fields, power diffuse radio emission (halos, relics, and mini-halos) \citep[e.g.][]{zuhone2016, vanweeren2019}. Setting aside projection effects for a moment, a substantial overlap between clusters hosting X-ray shocks/cold fronts and those with diffuse radio emission could be expected. Indeed, $\sim88\%$ of the clusters with at least one edge also host a radio halo, relic, or mini-halo. The converse is not symmetric: only about half of the clusters with diffuse radio emission show an edge in our X-ray analysis. This asymmetry could be naturally explained by projection and instrumental effects. Radio halos and mini-halos are largely orientation-independent, whereas simulations indicate that cold fronts are detectable with XMM only when their interface lies within $\sim30^\circ$ of the plane of the sky \citep{Ghizzardi2010}. Similar, if not more stringent, geometrical constraints apply to shock fronts in X-rays. This imposes a narrow detectability window.

\begin{figure}
    \centering
    \includegraphics[scale=0.96]{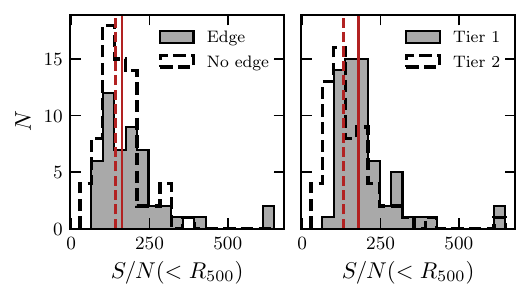}
    \caption{Distributions of the S/N measured within $R_{500}$ for clusters with and without edges (left panel), and Tier 1 and Tier 2 objects (right panel). The vertical lines show the median values for the different cluster populations.}
    \label{fig:S/N}
\end{figure}

Another interesting result of our analysis is the higher fraction of clusters hosting edges in the Tier 2 subsample compared to the Tier 1 (see Subsect. \ref{Sect:catalog_edges}). To investigate this behavior, we first assessed the role of the S/N (defined as $(C_s - C_b)/\sqrt{C_s+C_b}$, with $C_s$ being the source counts and $C_b$ the background counts) in the detection of edges. In Fig. \ref{fig:S/N}, we show the distribution of the S/N computed within $R_{500}$ for clusters with and without edges (left panel) and for clusters belonging to the Tier 1 or Tier 2 samples (right panel).
In the first case, the two distributions are similar, suggesting that the presence of an edge is not strongly driven by the overall data quality. In the second case, we found that Tier 2 clusters have lower S/N than Tier 1 objects, with median values of S/N$\sim 130$ and S/N$\sim 180$, respectively, and p-value of $0.001$. This suggests that the higher edge fraction observed in the Tier 2 is unlikely to be driven only by data quality. Moreover, previous morphological analysis of the CHEX-MATE clusters \citep{Campitiello2022} showed that Tier 1 and Tier 2 objects are not characterized by significantly different dynamical states, suggesting that the higher fraction of edges in the Tier 2, is not simply due to a higher fraction of disturbed objects. Taken together, these results may suggest a genuinely higher occurrence of detectable edges in massive and/or high-redshift clusters. However, for construction, mass and redshift are intrinsically coupled in CHEX-MATE, and the relative contribution of these two quantities cannot be disentangled with the present dataset. Addressing this issue will require larger cluster samples and comparisons with cosmological simulations.

With these considerations in mind, we can now attempt to interpret the results of our analysis. The fraction of objects exhibiting 0, or multiple (i.e. $\ge 1$) edges remains consistent across the different dynamical state of the clusters. However, the analysis presented in Sect. \ref{Subsect:comparison_Chandra}, shows that relaxed systems are preferentially affected by incompleteness. For this reason, the apparent similarity in the occurrence of edges observed in relaxed, mixed and disturbed systems should be interpreted with caution, as the true fraction of relaxed systems hosting an edge may be higher than observed. Moreover, most of the missed detections in relaxed systems correspond to centrally located or weak edges. For this reason, the most significant effect of incompleteness is likely an underestimation of the true occurrence of edges in relaxed systems, rather than a modification of the main trends observed for this population, namely lower $\mathcal{J}$ and $d_x$ values compared to the mixed and disturbed classes. We investigated how the $\mathcal{J}$ values of the recovered and not recovered edges change between the relaxed, mixed and disturbed classes. By comparing their medians, we found that relaxed and mixed recovered edges tend to have higher medians ($\mathcal{J}_R = 1.67$ and $\mathcal{J}_M = 1.72$, respectively) than the not recovered ones ($\mathcal{J}_R = 1.57$ and $\mathcal{J}_M = 1.55$, respectively). In disturbed systems instead the median values are the same ($\mathcal{J}_D = 2.0$). These results suggest that our conclusions regarding the properties of the detected edges are robust: relaxed edges remain preferentially weaker and closer to the cluster core than those found in disturbed systems.

Focusing on the properties of the detected edges, one of the most intriguing results of this work is the strong correlation observed between the BCG offset and the distance of the edge from the X-ray peak. As shown in Fig. \ref{fig:enter-label}, this trend appears to be driven primarily by the mixed and disturbed systems, while no clear correlation is visible among the relaxed clusters. Moreover, this is the only significant correlation found between an edge property and the morphological parameters of the host cluster. This behavior could be related to the different nature of these indicators: parameters such as $c$ and $w$ mainly describe the global properties of the ICM distribution, while the BCG offset directly measures the displacement between the peak of the X-ray emitting plasma and the brightest cluster galaxy, which is expected to trace the center of the underlying dark matter potential. The observed correlation therefore suggests more than a simple dependence on the overall disturbance level of the system. A possible interpretation is that both $d_X$ and the BCG-offset are driven by the same physical mechanism. During a merger, the perturbation can displace the ICM from the bottom of the gravitational potential. The perturbation responsible for this decoupling may therefore be the same mechanism that drives and compresses the ICM along the merger axis, leading to the formation of cold fronts and other edges. In this scenario, stronger perturbations would naturally produce both larger displacements between the X-ray peak and the BCG and edges located at larger distances from the cluster center. Compressive and intermittent turbulence, which can be driven or enhanced by merger-induced gas motions, may also contribute to this picture by generating substantial density fluctuations and projected edge-like SB structures \citep{Gaspari2013,Gaspari2014}. This interpretation is further supported by the marginal evidence (p-value $\sim0.06$) that edges in mixed and disturbed systems tend to exhibit lower values of $\mathcal{O}$, and are therefore more closely aligned with the direction connecting the BCG and the X-ray peak. If confirmed with larger samples, this behavior would be consistent with a picture in which edges preferentially develop along the direction of the merger-induced displacement of the gas with respect to the center of the gravitational potential. Moreover, the fact that the BCG offset--$d_X$ correlation is primarily observed in mixed and disturbed systems is broadly consistent with this scenario. These classes of objects are expected to be more strongly affected by merger-induced gas motions, with edges generated either by shock waves or by contact edges between ICM components at different temperatures during the merger process. In contrast, relaxed systems are expected to host predominantly sloshing cold fronts, generated by the oscillation of the ICM within the cluster potential following past minor or off-axis merger events. Such edges may develop along less predictable directions and are therefore not necessarily expected to follow the same BCG offset--$d_X$ relation. While dedicated comparisons with hydrodynamical simulations will be required to test this interpretation quantitatively, the observed trend points toward a possible physical connection between the displacement of the central gas distribution and the characteristic scale of the resulting edges.

To compare our results with edges reported in the literature, we considered the work by \cite{Ghizzardi2010}. Although, as we will explain below, there are significant differences between the two analyses, we chose as a first step to compare our results with those of \cite{Ghizzardi2010}, as this is the only one aimed to detect edges in a sample of galaxy clusters using XMM maps and, therefore, likely subject to the same limitations as our work. Their initial sample comprised 45 clusters with redshifts below 0.2 and fluxes of $f_x>1.7 \times 10^{-11}$ erg cm$^{-2}$ s$^{-1}$ in the 2–10 keV energy band. While this sample was considered representative of the cluster population, it was not complete. Due to the limited angular resolution, their final analysis focused on a subsample of 32 clusters at $z<0.075$. \cite{Ghizzardi2010} distinguished cold fronts associated with mergers, identified in systems showing clear merger morphology, often near the plane of the sky, from those associated with sloshing, found in otherwise relaxed clusters and attributed to minor or off-axis perturbations. A similar dichotomy has been adopted elsewhere; for example, \citet{Owers2009} selected clusters by sharp surface-brightness edges in Chandra images and identified both fronts embedded in obviously merging systems and fronts in apparently relaxed systems where the cold front is the primary merger signature. Focusing on sloshing cold fronts in \citet{Ghizzardi2010}, 10 of 23 relaxed/non-merging clusters ($\sim43\%$) host a cold front, 9 of which are cool cores. A similar fraction is found in CHEX–MATE: restricting likewise to relaxed (non-merging) systems, 5 of 15 clusters ($\sim33\%$) host at least one cold front. However, as mentioned above, the fraction of edges in relaxed systems is likely underestimated in our analysis, implying that it should likely be regarded as a lower limit. For additional context, \citet{Markevitch2003} analyzed 37 relaxed clusters with Chandra and found cold fronts in about two thirds of the sample; among 18 cool‐core clusters, $\approx80\%$ showed a central front separating cool core gas from hotter ambient ICM. They argued that cold fronts are likely common in relaxed systems, and can be triggered not only by minor/off-axis mergers but also by AGN outbursts \citep{Quilis2001}.

\section{Summary and conclusions}\label{Sect:conclusion}

In this paper, we have examined the ICM distribution of the 116 CHEX-MATE galaxy clusters to detect local edges in the X-ray surface brightness maps, likely imprinted by shocks and cold fronts. We summarize our main results below:
\begin{itemize}

    \item We have implemented, and validated, a method for the detection of edges in the X-ray (soft) images of galaxy clusters observed by XMM (see Sect.~\ref{Sect:method_edges}). 
    This allowed us to perform, for the first time, a systematic analysis of such edges on an SZ-selected, and homogeneously X-ray exposed in the region across $R_{500}$, sample extending up to redshift $z \sim 0.6$.

    \item From a comparison with the detections available in the literature, we found that our XMM exposures, tailored to estimate the spectral temperature in the annulus [0.8-1.2] $R_{500}$ with $\sim$15\% statistical uncertainty at $1 \sigma$, are able to retrieve only $\sim$ half of the edges reported in the literature. We also find that the incompleteness mainly affects weak, central cold fronts in relaxed systems (see Sect.~\ref{Subsect:comparison_Chandra}).

    \item By applying this method to the whole sample,  we detected 66 edges, 32 of which are newly detected, in 48 out of 116 ($\sim 41$\%) CHEX-MATE objects. This corresponds to the fraction of clusters hosting edges that are detectable with the present XMM data and our analysis procedure.
    
    \item We found that (see Sect.~\ref{Sect:catalog_edges}): 
    no particular difference is observed among the occurrence of edges in relaxed, disturbed, and mixed systems; the fraction of edges is significantly higher at high masses (i.e. $ M_{500} > 7 \cdot 10^{14} \text{M}_{\odot}$, $\sim 54\%$) and at high redshift (i.e. $z>0.2$, $\sim50\%$). 

    \item We characterized each edge by measuring four parameters. These are: the magnitude of the jump, $\mathcal{J}$, the distance from the X-ray peak $d_x$, the angular extension, $\theta$, and the orientation with the ellipses that better approximates the X-ray emission of the cluster $\mathcal{O}$ (see Sect. \ref{Sect:statistical_edge} for more details on their definition). We found that edges in disturbed systems are usually stronger and farther from the X-ray peak than those in relaxed systems. By comparing the $\mathcal{O}$ median values, we also observed that edges in disturbed systems seem more aligned than edges in relaxed and mixed systems.
    
    \item We checked whether clusters with or without edges show different distributions of the morphological parameters presented in \cite{Campitiello2022}, and of the BCG-offset. Using a KS test, we found no significant difference between the two populations of objects. Furthermore, we noticed a strong correlation between the BCG-offset and the distance of the edge from the X-ray peak (see Subsect. \ref{Subsect:morpho_param_edges}), which is especially driven by the mixed and disturbed objects. This behavior is consistent with mergers simultaneously displacing the BCG and generating large-scale gas motions that give rise to X-ray edges.

    \item We used curvelet temperature maps to classify edges as shocks or cold fronts. We found that: 8 possible shocks (about 20\% of the edges that we are able to characterize spectroscopically) and 34 possible cold fronts are present in our sample. The fractions of clusters hosting at least one cold front are comparable in mixed and disturbed systems ($\sim$20\% and $\sim$19\%), while shocks are rare in both classes ($\sim$7\% and $\sim$4\%, respectively). No shock candidates are identified in relaxed systems. Using the KS tests, we compared these two populations of objects, finding that shocks tend to be located at greater distances from the X-ray peak and to be less aligned with the global morphology of the X-ray emission (see Subsect.~\ref{Subsect:classification}).
    
    \item By examining the link with the radio emission, we found that in a total of 79 CHEX-MATE objects, covered by radio observations and hosting a diffuse radio source (i.e., a radio halo, a radio relic or both), 38 show an edge (48\%, see Subsect. \ref{Subsect:radio_emission_edges}).
    
    \item We compared our results with previous studies. We found that the fraction of relaxed clusters hosting a cold front ($\sim 33 \%$) is close to the fraction of sloshing cold fronts found in \cite{Ghizzardi2010} ($\sim 43 \%$).
\end{itemize}

A more detailed physical interpretation of the detected structures will be addressed in future work. This will require the development of toy models—either analytical or based on hydrodynamic simulations—to establish a more direct association between the observed features and the underlying physical mechanisms that generate them. Additionally, simulations could help better assess the systematic effects affecting detection—such as signal-to-noise ratio (linked to exposure time), position relative to the cluster center, and CCD placement—that will be necessary to refine our understanding. \\

\noindent \textbf{Data availability.} Appendices A--I are available on \href{https://doi.org/10.5281/zenodo.22715797}{Zenodo}.
\begin{acknowledgements}
Work at Argonne National Lab is supported by UChicago Argonne LLC, Operator of Argonne National Laboratory (Argonne). Argonne, a U.S. Department of Energy Office of Science Laboratory, is operated under contract no. DE-AC02-06CH11357.
We acknowledge the financial contribution from the contracts
Prin-MUR 2022 supported by Next Generation EU (M4.C2.1.1, n.20227RNLY3 The concordance cosmological model: stress-tests with galaxy clusters), and from the European Union’s Horizon 2020 Programme under the AHEAD2020 project (grant agreement n. 871158).
This research was supported by the International Space Science Institute (ISSI) in Bern, through ISSI International Team project \#565 (Multi-Wavelength Studies of the Culmination of Structure Formation in the Universe).
LL acknowledges the financial contribution from the INAF grant 1.05.12.04.01. M.G. acknowledges support from the ERC Consolidator Grant BlackHoleWeather (101086804). WF acknowledges support from the Smithsonian Institution, the Chandra High Resolution Camera Project through NASA contract NAS8-0306, NASA Grant 80NSSC19K0116 and Chandra Grant GO1-22132X. BJM acknowledges support from Science and Technology Facilities Council grants  ST/V000454/1 and ST/Y002008/1. GWP acknowledges long-term support from CNES, the French space agency.
      
\end{acknowledgements}

\bibliographystyle{aa}
\bibliography{biblio}

@ARTICLE{Akamatsu2017,
       author = {{Akamatsu}, H. and {Fujita}, Y. and {Akahori}, T. and {Ishisaki}, Y. and {Hayashida}, K. and {Hoshino}, A. and {Mernier}, F. and {Yoshikawa}, K. and {Sato}, K. and {Kaastra}, J.~S.},
        title = "{Properties of the cosmological filament between two clusters: A possible detection of a large-scale accretion shock by Suzaku}",
      journal = {\aap},
         year = 2017,
        month = sep,
       volume = {606},
          eid = {A1},
        pages = {A1},
          doi = {10.1051/0004-6361/201730497},
archivePrefix = {arXiv},
       eprint = {1704.05843},
 primaryClass = {astro-ph.HE},
       adsurl = {https://ui.adsabs.harvard.edu/abs/2017A\&A...606A...1A}
}

@ARTICLE{Bartalucci23,
       author = {{Bartalucci}, I. and {Molendi}, S. and {Rasia}, E. and {Pratt}, G.~W. and {Arnaud}, M. and {Rossetti}, M. and {Gastaldello}, F. and {Eckert}, D. and {Balboni}, M. and {Borgani}, S. and {Bourdin}, H. and {Campitiello}, M.~G. and {De Grandi}, S. and {De Petris}, M. and {Duffy}, R.~T. and {Ettori}, S. and {Ferragamo}, A. and {Gaspari}, M. and {Gavazzi}, R. and {Ghizzardi}, S. and {Iqbal}, A. and {Kay}, S.~T. and {Lovisari}, L. and {Mazzotta}, P. and {Maughan}, B.~J. and {Pointecouteau}, E. and {Riva}, G. and {Sereno}, M.},
        title = "{CHEX-MATE: Constraining the origin of the scatter in galaxy cluster radial X-ray surface brightness profiles}",
      journal = {A\&A},
         year = 2023,
        month = may,
       volume = {674},
          eid = {A179},
        pages = {A179},
          doi = {10.1051/0004-6361/202346189},
archivePrefix = {arXiv},
       eprint = {2305.03082},
 primaryClass = {astro-ph.CO},
       adsurl = {https://ui.adsabs.harvard.edu/abs/2023A\&A...674A.179B}
}

@ARTICLE{Birzan2004,
       author = {{B{\^\i}rzan}, L. and {Rafferty}, D.~A. and {McNamara}, B.~R. and {Wise}, M.~W. and {Nulsen}, P.~E.~J.},
        title = "{A Systematic Study of Radio-induced X-Ray Cavities in Clusters, Groups, and Galaxies}",
      journal = {\apj},
         year = 2004,
        month = jun,
       volume = {607},
       number = {2},
        pages = {800-809},
          doi = {10.1086/383519},
archivePrefix = {arXiv},
       eprint = {astro-ph/0402348},
 primaryClass = {astro-ph},
       adsurl = {https://ui.adsabs.harvard.edu/abs/2004ApJ...607..800B}
}

@ARTICLE{Botteon2016,
       author = {{Botteon}, A. and {Gastaldello}, F. and {Brunetti}, G. and {Dallacasa}, D.},
        title = "{A shock at the radio relic position in Abell 115}",
      journal = {\mnras},
         year = 2016,
        month = jul,
       volume = {460},
       number = {1},
        pages = {L84-L88},
          doi = {10.1093/mnrasl/slw082},
archivePrefix = {arXiv},
       eprint = {1604.07823},
 primaryClass = {astro-ph.HE},
       adsurl = {https://ui.adsabs.harvard.edu/abs/2016MNRAS.460L..84B}
}

@ARTICLE{Botteon2016b,
       author = {{Botteon}, A. and {Gastaldello}, F. and {Brunetti}, G. and {Kale}, R.},
        title = "{A M {\ensuremath{\gtrsim}} 3 shock in `El Gordo' cluster and the origin of the radio relic}",
      journal = {\mnras},
         year = 2016,
        month = dec,
       volume = {463},
       number = {2},
        pages = {1534-1542},
          doi = {10.1093/mnras/stw2089},
archivePrefix = {arXiv},
       eprint = {1607.04641},
 primaryClass = {astro-ph.HE},
       adsurl = {https://ui.adsabs.harvard.edu/abs/2016MNRAS.463.1534B}
}

@ARTICLE{Botteon2018,
       author = {{Botteon}, A. and {Gastaldello}, F. and {Brunetti}, G.},
        title = "{Shocks and cold fronts in merging and massive galaxy clusters: new detections with Chandra}",
      journal = {\mnras},
         year = 2018,
        month = jun,
       volume = {476},
       number = {4},
        pages = {5591-5620},
          doi = {10.1093/mnras/sty598},
archivePrefix = {arXiv},
       eprint = {1707.07038},
 primaryClass = {astro-ph.HE},
       adsurl = {https://ui.adsabs.harvard.edu/abs/2018MNRAS.476.5591B}
}

@ARTICLE{Botteon2023,
       author = {{Botteon}, Andrea and {Markevitch}, Maxim and {van Weeren}, Reinout J. and {Brunetti}, Gianfranco and {Shimwell}, Timothy W.},
        title = "{Surface brightness discontinuities in radio halos. Insights from the MeerKAT Galaxy Cluster Legacy Survey}",
      journal = {\aap},
         year = 2023,
        month = jun,
       volume = {674},
          eid = {A53},
        pages = {A53},
          doi = {10.1051/0004-6361/202346150},
archivePrefix = {arXiv},
       eprint = {2302.07881},
 primaryClass = {astro-ph.CO},
       adsurl = {https://ui.adsabs.harvard.edu/abs/2023A\&A...674A..53B}
}

@ARTICLE{Bourdin2004,
       author = {{Bourdin}, H. and {Sauvageot}, J.-L. and {Slezak}, E. and {Bijaoui}, A. and {Teyssier}, R.},
        title = "{Temperature map computation for X-ray clusters of galaxies}",
      journal = {\aap},
         year = 2004,
        month = feb,
       volume = {414},
        pages = {429-443},
          doi = {10.1051/0004-6361:20031662},
       adsurl = {https://ui.adsabs.harvard.edu/abs/2004A\&A...414..429B}
}

@ARTICLE{Bourdin2008,
       author = {{Bourdin}, H. and {Mazzotta}, P.},
        title = "{Temperature structure of the intergalactic medium within seven nearby and bright clusters of galaxies observed with XMM-Newton}",
      journal = {\aap},
         year = 2008,
        month = feb,
       volume = {479},
       number = {2},
        pages = {307-320},
          doi = {10.1051/0004-6361:20065758},
archivePrefix = {arXiv},
       eprint = {0802.1866},
 primaryClass = {astro-ph},
       adsurl = {https://ui.adsabs.harvard.edu/abs/2008A\&A...479..307B}
}

@ARTICLE{Bourdin2013,
       author = {{Bourdin}, H. and {Mazzotta}, P. and {Markevitch}, M. and {Giacintucci}, S. and {Brunetti}, G.},
        title = "{Shock Heating of the Merging Galaxy Cluster A521}",
      journal = {\apj},
         year = 2013,
        month = feb,
       volume = {764},
       number = {1},
          eid = {82},
        pages = {82},
          doi = {10.1088/0004-637X/764/1/82},
archivePrefix = {arXiv},
       eprint = {1302.0696},
 primaryClass = {astro-ph.CO},
       adsurl = {https://ui.adsabs.harvard.edu/abs/2013ApJ...764...82B}
}

@ARTICLE{Bourdin2015,
       author = {{Bourdin}, H. and {Mazzotta}, P. and {Rasia}, E.},
        title = "{Spectral Imaging of Galaxy Clusters with Planck}",
      journal = {\apj},
         year = 2015,
        month = dec,
       volume = {815},
       number = {2},
          eid = {92},
        pages = {92},
          doi = {10.1088/0004-637X/815/2/92},
archivePrefix = {arXiv},
       eprint = {1601.06323},
 primaryClass = {astro-ph.CO},
       adsurl = {https://ui.adsabs.harvard.edu/abs/2015ApJ...815...92B}
}

@ARTICLE{Brown2011,
       author = {{Brown}, Shea and {Rudnick}, Lawrence},
        title = "{Diffuse radio emission in/around the Coma cluster: beyond simple accretion}",
      journal = {\mnras},
         year = 2011,
        month = mar,
       volume = {412},
       number = {1},
        pages = {2-12},
          doi = {10.1111/j.1365-2966.2010.17738.x},
archivePrefix = {arXiv},
       eprint = {1009.4258},
 primaryClass = {astro-ph.CO},
       adsurl = {https://ui.adsabs.harvard.edu/abs/2011MNRAS.412....2B}
}

@ARTICLE{Campitiello2022,
       author = {{Campitiello}, M.~G. and {Ettori}, S. and {Lovisari}, L. and {Bartalucci}, I. and {Eckert}, D. and {Rasia}, E. and {Rossetti}, M. and {Gastaldello}, F. and {Pratt}, G.~W. and {Maughan}, B. and {Pointecouteau}, E. and {Sereno}, M. and {Biffi}, V. and {Borgani}, S. and {De Luca}, F. and {De Petris}, M. and {Gaspari}, M. and {Ghizzardi}, S. and {Mazzotta}, P. and {Molendi}, S.},
        title = "{CHEX-MATE: Morphological analysis of the sample}",
      journal = {\aap},
         year = 2022,
        month = sep,
       volume = {665},
          eid = {A117},
        pages = {A117},
          doi = {10.1051/0004-6361/202243470},
archivePrefix = {arXiv},
       eprint = {2205.11326},
 primaryClass = {astro-ph.CO},
       adsurl = {https://ui.adsabs.harvard.edu/abs/2022A\&A...665A.117C}
}

@article{Canny1986,
  author = {Canny, John},
  title = {A Computational Approach to Edge Detection},
  journal = {IEEE Transactions on Pattern Analysis and Machine Intelligence},
  volume = {8},
  number = {6},
  pages = {679--698},
  year = {1986}
}

@ARTICLE{Cavaliere1976,
       author = {{Cavaliere}, A. and {Fusco-Femiano}, R.},
        title = "{X-rays from hot plasma in clusters of galaxies.}",
      journal = {\aap},
         year = 1976,
        month = may,
       volume = {49},
        pages = {137-144},
       adsurl = {https://ui.adsabs.harvard.edu/abs/1976A\&A....49..137C}
}

@ARTICLE{chexmate,
       author = {{CHEX-MATE Collaboration} and {Arnaud}, M. and {Ettori}, S. and {Pratt}, G.~W. and {Rossetti}, M. and {Eckert}, D. and {Gastaldello}, F. and {Gavazzi}, R. and {Kay}, S.~T. and {Lovisari}, L. and {Maughan}, B.~J. and {Pointecouteau}, E. and {Sereno}, M. and {Bartalucci}, I. and {Bonafede}, A. and {Bourdin}, H. and {Cassano}, R. and {Duffy}, R.~T. and {Iqbal}, A. and {Maurogordato}, S. and {Rasia}, E. and {Sayers}, J. and {Andrade-Santos}, F. and {Aussel}, H. and {Barnes}, D.~J. and {Barrena}, R. and {Borgani}, S. and {Burkutean}, S. and {Clerc}, N. and {Corasaniti}, P. -S. and {Cuillandre}, J. -C. and {De Grandi}, S. and {De Petris}, M. and {Dolag}, K. and {Donahue}, M. and {Ferragamo}, A. and {Gaspari}, M. and {Ghizzardi}, S. and {Gitti}, M. and {Haines}, C.~P. and {Jauzac}, M. and {Johnston-Hollitt}, M. and {Jones}, C. and {K{\'e}ruzor{\'e}}, F. and {LeBrun}, A.~M.~C. and {Mayet}, F. and {Mazzotta}, P. and {Melin}, J. -B. and {Molendi}, S. and {Nonino}, M. and {Okabe}, N. and {Paltani}, S. and {Perotto}, L. and {Pires}, S. and {Radovich}, M. and {Rubino-Martin}, J. -A. and {Salvati}, L. and {Saro}, A. and {Sartoris}, B. and {Schellenberger}, G. and {Streblyanska}, A. and {Tarr{\'\i}o}, P. and {Tozzi}, P. and {Umetsu}, K. and {van der Burg}, R.~F.~J. and {Vazza}, F. and {Venturi}, T. and {Yepes}, G. and {Zarattini}, S.},
        title = "{The Cluster HEritage project with XMM-Newton: Mass Assembly and Thermodynamics at the Endpoint of structure formation. I. Programme overview}",
      journal = {\aap},
         year = 2021,
        month = jun,
       volume = {650},
          eid = {A104},
        pages = {A104},
          doi = {10.1051/0004-6361/202039632},
       adsurl = {https://ui.adsabs.harvard.edu/abs/2021A\&A...650A.104C}
}

@ARTICLE{Choudhury2025,
       author = {{Choudhury}, Prakriti Pal and {Reynolds}, Christopher S.},
        title = "{Cold fronts in galaxy clusters - I. A case for the large-scale global eigenmodes in unmagnetized and weakly magnetized cluster core}",
      journal = {\mnras},
         year = 2025,
        month = mar,
       volume = {537},
       number = {4},
        pages = {3194-3209},
          doi = {10.1093/mnras/staf184},
archivePrefix = {arXiv},
       eprint = {2408.03988},
 primaryClass = {astro-ph.GA},
       adsurl = {https://ui.adsabs.harvard.edu/abs/2025MNRAS.537.3194C}
}

@ARTICLE{Churazov2000,
       author = {{Churazov}, E. and {Forman}, W. and {Jones}, C. and {B{\"o}hringer}, H.},
        title = "{Asymmetric, arc minute scale structures around NGC 1275}",
      journal = {\aap},
         year = 2000,
        month = apr,
       volume = {356},
        pages = {788-794},
          doi = {10.48550/arXiv.astro-ph/0002375},
archivePrefix = {arXiv},
       eprint = {astro-ph/0002375},
 primaryClass = {astro-ph},
       adsurl = {https://ui.adsabs.harvard.edu/abs/2000A&A...356..788C}
}

@ARTICLE{Churazov2001,
       author = {{Churazov}, E. and {Br{\"u}ggen}, M. and {Kaiser}, C.~R. and {B{\"o}hringer}, H. and {Forman}, W.},
        title = "{Evolution of Buoyant Bubbles in M87}",
      journal = {\apj},
         year = 2001,
        month = jun,
       volume = {554},
       number = {1},
        pages = {261-273},
          doi = {10.1086/321357},
archivePrefix = {arXiv},
       eprint = {astro-ph/0008215},
 primaryClass = {astro-ph},
       adsurl = {https://ui.adsabs.harvard.edu/abs/2001ApJ...554..261C}
}

@ARTICLE{Eckert2016,
       author = {{Eckert}, D. and {Jauzac}, M. and {Vazza}, F. and {Owers}, M.~S. and {Kneib}, J. -P. and {Tchernin}, C. and {Intema}, H. and {Knowles}, K.},
        title = "{A shock front at the radio relic of Abell 2744}",
      journal = {\mnras},
         year = 2016,
        month = sep,
       volume = {461},
       number = {2},
        pages = {1302-1307},
          doi = {10.1093/mnras/stw1435},
archivePrefix = {arXiv},
       eprint = {1603.02272},
 primaryClass = {astro-ph.HE},
       adsurl = {https://ui.adsabs.harvard.edu/abs/2016MNRAS.461.1302E}
}

@ARTICLE{Eckert2020,
       author = {{Eckert}, Dominique and {Finoguenov}, Alexis and {Ghirardini}, Vittorio and {Grandis}, Sebastian and {Kaefer}, Florian and {Sanders}, Jeremy and {Ramos-Ceja}, Miriam},
        title = "{Low-scatter galaxy cluster mass proxies for the eROSITA all-sky survey}",
      journal = {The Open Journal of Astrophysics},
         year = 2020,
        month = sep,
       volume = {3},
          eid = {12},
        pages = {12},
          doi = {10.21105/astro.2009.13944},
archivePrefix = {arXiv},
       eprint = {2009.03944},
 primaryClass = {astro-ph.CO},
       adsurl = {https://ui.adsabs.harvard.edu/abs/2020OJAp....3E..12E}
}

@ARTICLE{Ettori2010,
       author = {{Ettori}, S. and {Gastaldello}, F. and {Leccardi}, A. and {Molendi}, S. and {Rossetti}, M. and {Buote}, D. and {Meneghetti}, M.},
        title = "{Mass profiles and c-M$_{DM}$ relation in X-ray luminous galaxy clusters}",
      journal = {\aap},
         year = 2010,
        month = dec,
       volume = {524},
          eid = {A68},
        pages = {A68},
          doi = {10.1051/0004-6361/201015271},
       adsurl = {https://ui.adsabs.harvard.edu/abs/2010A\&A...524A..68E}
}

@ARTICLE{Ettori2013,
       author = {{Ettori}, S. and {Gastaldello}, F. and {Gitti}, M. and {O'Sullivan}, E. and {Gaspari}, M. and {Brighenti}, F. and {David}, L. and {Edge}, A.~C.},
        title = "{Cold fronts and metal anisotropies in the X-ray cool core of the galaxy cluster Zw 1742+3306}",
      journal = {\aap},
         year = 2013,
        month = jul,
       volume = {555},
          eid = {A93},
        pages = {A93},
          doi = {10.1051/0004-6361/201321107},
archivePrefix = {arXiv},
       eprint = {1305.3926},
 primaryClass = {astro-ph.CO},
       adsurl = {https://ui.adsabs.harvard.edu/abs/2013A\&A...555A..93E}
}

@ARTICLE{Fabian2006,
       author = {{Fabian}, A.~C. and {Sanders}, J.~S. and {Taylor}, G.~B. and {Allen}, S.~W. and {Crawford}, C.~S. and {Johnstone}, R.~M. and {Iwasawa}, K.},
        title = "{A very deep Chandra observation of the Perseus cluster: shocks, ripples and conduction}",
      journal = {\mnras},
         year = 2006,
        month = feb,
       volume = {366},
       number = {2},
        pages = {417-428},
          doi = {10.1111/j.1365-2966.2005.09896.x},
archivePrefix = {arXiv},
       eprint = {astro-ph/0510476},
 primaryClass = {astro-ph},
       adsurl = {https://ui.adsabs.harvard.edu/abs/2006MNRAS.366..417F}
}

@ARTICLE{Gaspari2011,
       author = {{Gaspari}, M. and {Melioli}, C. and {Brighenti}, F. and {D'Ercole}, A.},
        title = "{The dance of heating and cooling in galaxy clusters: three-dimensional simulations of self-regulated active galactic nuclei outflows}",
      journal = {\mnras},
         year = 2011,
        month = feb,
       volume = {411},
       number = {1},
        pages = {349-372},
          doi = {10.1111/j.1365-2966.2010.17688.x},
archivePrefix = {arXiv},
       eprint = {1007.0674},
 primaryClass = {astro-ph.CO},
       adsurl = {https://ui.adsabs.harvard.edu/abs/2011MNRAS.411..349G}
}

@ARTICLE{Gaspari2013,
       author = {{Gaspari}, M. and {Churazov}, E.},
        title = "{Constraining turbulence and conduction in the hot ICM through density perturbations}",
      journal = {\aap},
         year = 2013,
        month = nov,
       volume = {559},
          eid = {A78},
        pages = {A78},
          doi = {10.1051/0004-6361/201322295},
archivePrefix = {arXiv},
       eprint = {1307.4397},
 primaryClass = {astro-ph.CO},
       adsurl = {https://ui.adsabs.harvard.edu/abs/2013A&A...559A..78G}
}

@ARTICLE{Gaspari2014,
       author = {{Gaspari}, M. and {Churazov}, E. and {Nagai}, D. and {Lau}, E.~T. and {Zhuravleva}, I.},
        title = "{The relation between gas density and velocity power spectra in galaxy clusters: High-resolution hydrodynamic simulations and the role of conduction}",
      journal = {\aap},
         year = 2014,
        month = sep,
       volume = {569},
          eid = {A67},
        pages = {A67},
          doi = {10.1051/0004-6361/201424043},
archivePrefix = {arXiv},
       eprint = {1404.5302},
 primaryClass = {astro-ph.CO},
       adsurl = {https://ui.adsabs.harvard.edu/abs/2014A\&A...569A..67G}
}

@ARTICLE{Ghirardini2019,
       author = {{Ghirardini}, V. and {Eckert}, D. and {Ettori}, S. and {Pointecouteau}, E. and {Molendi}, S. and {Gaspari}, M. and {Rossetti}, M. and {De Grandi}, S. and {Roncarelli}, M. and {Bourdin}, H. and {Mazzotta}, P. and {Rasia}, E. and {Vazza}, F.},
        title = "{Universal thermodynamic properties of the intracluster medium over two decades in radius in the X-COP sample}",
      journal = {\aap},
         year = 2019,
        month = jan,
       volume = {621},
          eid = {A41},
        pages = {A41},
          doi = {10.1051/0004-6361/201833325},
archivePrefix = {arXiv},
       eprint = {1805.00042},
 primaryClass = {astro-ph.CO},
       adsurl = {https://ui.adsabs.harvard.edu/abs/2019A\&A...621A..41G}
}

@ARTICLE{Giacintucci2008,
       author = {{Giacintucci}, S. and {Venturi}, T. and {Macario}, G. and {Dallacasa}, D. and {Brunetti}, G. and {Markevitch}, M. and {Cassano}, R. and {Bardelli}, S. and {Athreya}, R.},
        title = "{Shock acceleration as origin of the radio relic in A 521?}",
      journal = {\aap},
         year = 2008,
        month = aug,
       volume = {486},
       number = {2},
        pages = {347-358},
          doi = {10.1051/0004-6361:200809459},
archivePrefix = {arXiv},
       eprint = {0803.4127},
 primaryClass = {astro-ph},
       adsurl = {https://ui.adsabs.harvard.edu/abs/2008A\&A...486..347G}
}

@ARTICLE{Giacintucci2014,
       author = {{Giacintucci}, S. and {Markevitch}, M. and {Brunetti}, G. and {ZuHone}, J.~A. and {Venturi}, T. and {Mazzotta}, P. and {Bourdin}, H.},
        title = "{Mapping the Particle Acceleration in the Cool Core of the Galaxy Cluster RX J1720.1+2638}",
      journal = {\apj},
         year = 2014,
        month = nov,
       volume = {795},
       number = {1},
          eid = {73},
        pages = {73},
          doi = {10.1088/0004-637X/795/1/73},
archivePrefix = {arXiv},
       eprint = {1403.2820},
 primaryClass = {astro-ph.CO},
       adsurl = {https://ui.adsabs.harvard.edu/abs/2014ApJ...795...73G}
}

@ARTICLE{Giacintucci2014b,
       author = {{Giacintucci}, Simona and {Markevitch}, Maxim and {Venturi}, Tiziana and {Clarke}, Tracy E. and {Cassano}, Rossella and {Mazzotta}, Pasquale},
        title = "{New Detections of Radio Minihalos in Cool Cores of Galaxy Clusters}",
      journal = {\apj},
         year = 2014,
        month = jan,
       volume = {781},
       number = {1},
          eid = {9},
        pages = {9},
          doi = {10.1088/0004-637X/781/1/9},
archivePrefix = {arXiv},
       eprint = {1311.5248},
 primaryClass = {astro-ph.CO},
       adsurl = {https://ui.adsabs.harvard.edu/abs/2014ApJ...781....9G}
}

@ARTICLE{Gitti2025,
       author = {{Gitti}, M. and {Bonafede}, A. and {Brighenti}, F. and {Ubertosi}, F. and {Balboni}, M. and {Gastaldello}, F. and {Botteon}, A. and {Forman}, W. and {van Weeren}, R.~J. and {Br{\"u}ggen}, M. and {Rajpurohit}, K. and {Jones}, C.},
        title = "{Deep Chandra observations of PLCKG287.0+32.9: a clear detection of a shock front in a heated former cool core}",
      journal = {A\&A},
         year = 2025,
        month = mar,
       volume = {697},
          eid = {A72},
        pages = {A72},
          doi = {10.1051/0004-6361/202453450},
archivePrefix = {arXiv},
       eprint = {2503.13735},
 primaryClass = {astro-ph.CO},
       adsurl = {https://ui.adsabs.harvard.edu/abs/2025A\&A...697A..72G}
}

@ARTICLE{hlavacek2012,
       author = {{Hlavacek-Larrondo}, J. and {Fabian}, A.~C. and {Edge}, A.~C. and {Ebeling}, H. and {Sanders}, J.~S. and {Hogan}, M.~T. and {Taylor}, G.~B.},
        title = "{Extreme AGN feedback in the MAssive Cluster Survey: a detailed study of X-ray cavities at z>0.3}",
      journal = {\mnras},
         year = 2012,
        month = apr,
       volume = {421},
       number = {2},
        pages = {1360-1384},
          doi = {10.1111/j.1365-2966.2011.20405.x},
archivePrefix = {arXiv},
       eprint = {1110.0489},
 primaryClass = {astro-ph.CO},
       adsurl = {https://ui.adsabs.harvard.edu/abs/2012MNRAS.421.1360H}
}

@ARTICLE{Ghizzardi2010,
       author = {{Ghizzardi}, S. and {Rossetti}, M. and {Molendi}, S.},
        title = "{Cold fronts in galaxy clusters}",
      journal = {\aap},
         year = 2010,
        month = jun,
       volume = {516},
          eid = {A32},
        pages = {A32},
          doi = {10.1051/0004-6361/200912496},
archivePrefix = {arXiv},
       eprint = {1003.1051},
 primaryClass = {astro-ph.CO},
       adsurl = {https://ui.adsabs.harvard.edu/abs/2010A\&A...516A..32G}
}

@ARTICLE{Hurier2019,
       author = {{Hurier}, G. and {Adam}, R. and {Keshet}, U.},
        title = "{First detection of a virial shock with SZ data: implication for the mass accretion rate of Abell 2319}",
      journal = {\aap},
         year = 2019,
        month = feb,
       volume = {622},
          eid = {A136},
        pages = {A136},
          doi = {10.1051/0004-6361/201732468},
archivePrefix = {arXiv},
       eprint = {1712.05762},
 primaryClass = {astro-ph.CO},
       adsurl = {https://ui.adsabs.harvard.edu/abs/2019A\&A...622A.136H}
}

@ARTICLE{Ichinohe2015,
       author = {{Ichinohe}, Y. and {Werner}, N. and {Simionescu}, A. and {Allen}, S.~W. and {Canning}, R.~E.~A. and {Ehlert}, S. and {Mernier}, F. and {Takahashi}, T.},
        title = "{The growth of the galaxy cluster Abell 85: mergers, shocks, stripping and seeding of clumping}",
      journal = {\mnras},
         year = 2015,
        month = apr,
       volume = {448},
       number = {3},
        pages = {2971-2986},
          doi = {10.1093/mnras/stv217},
archivePrefix = {arXiv},
       eprint = {1410.1955},
 primaryClass = {astro-ph.HE},
       adsurl = {https://ui.adsabs.harvard.edu/abs/2015MNRAS.448.2971I}
}

@ARTICLE{Ichinohe2017,
       author = {{Ichinohe}, Y. and {Simionescu}, A. and {Werner}, N. and {Takahashi}, T.},
        title = "{An azimuthally resolved study of the cold front in Abell 3667}",
      journal = {\mnras},
         year = 2017,
        month = may,
       volume = {467},
       number = {3},
        pages = {3662-3676},
          doi = {10.1093/mnras/stx280},
archivePrefix = {arXiv},
       eprint = {1702.01026},
 primaryClass = {astro-ph.HE},
       adsurl = {https://ui.adsabs.harvard.edu/abs/2017MNRAS.467.3662I}
}

@ARTICLE{Ichinohe2019,
       author = {{Ichinohe}, Y. and {Simionescu}, A. and {Werner}, N. and {Fabian}, A.~C. and {Takahashi}, T.},
        title = "{Substructures associated with the sloshing cold front in the Perseus cluster}",
      journal = {\mnras},
         year = 2019,
        month = feb,
       volume = {483},
       number = {2},
        pages = {1744-1753},
          doi = {10.1093/mnras/sty3257},
archivePrefix = {arXiv},
       eprint = {1810.07380},
 primaryClass = {astro-ph.HE},
       adsurl = {https://ui.adsabs.harvard.edu/abs/2019MNRAS.483.1744I}
}

@ARTICLE{Li2026,
       author = {{Li}, I-Hsuan and {Ueda}, Shutaro and {Chiu}, I-Non and {Umetsu}, Keiichi},
        title = "{An azimuthally resolved study of sloshing cold fronts in three nearby galaxy clusters}",
      journal = {arXiv e-prints},
         year = 2026,
        month = jan,
          eid = {arXiv:2601.14392},
        pages = {arXiv:2601.14392},
          doi = {10.48550/arXiv.2601.14392},
archivePrefix = {arXiv},
       eprint = {2601.14392},
 primaryClass = {astro-ph.CO},
       adsurl = {https://ui.adsabs.harvard.edu/abs/2026arXiv260114392L}
}

@ARTICLE{Liu2019,
       author = {{Liu}, Wenhao and {Sun}, Ming and {Nulsen}, Paul and {Clarke}, Tracy and {Sarazin}, Craig and {Forman}, William and {Gaspari}, Massimo and {Giacintucci}, Simona and {Lal}, Dharam Vir and {Edge}, Tim},
        title = "{AGN feedback in galaxy group 3C 88: cavities, shock, and jet reorientation}",
      journal = {\mnras},
         year = 2019,
        month = apr,
       volume = {484},
       number = {3},
        pages = {3376-3392},
          doi = {10.1093/mnras/stz229},
archivePrefix = {arXiv},
       eprint = {1806.04692},
 primaryClass = {astro-ph.HE},
       adsurl = {https://ui.adsabs.harvard.edu/abs/2019MNRAS.484.3376L}
}

@ARTICLE{Lyutikov2006,
       author = {{Lyutikov}, M.},
        title = "{Magnetic draping of merging cores and radio bubbles in clusters of galaxies}",
      journal = {\mnras},
         year = 2006,
        month = nov,
       volume = {373},
       number = {1},
        pages = {73-78},
          doi = {10.1111/j.1365-2966.2006.10835.x},
archivePrefix = {arXiv},
       eprint = {astro-ph/0604178},
 primaryClass = {astro-ph},
       adsurl = {https://ui.adsabs.harvard.edu/abs/2006MNRAS.373...73L}
}

@ARTICLE{Macario2011,
       author = {{Macario}, Giulia and {Markevitch}, Maxim and {Giacintucci}, Simona and {Brunetti}, Gianfranco and {Venturi}, Tiziana and {Murray}, Stephen S.},
        title = "{A Shock Front in the Merging Galaxy Cluster A754: X-ray and Radio Observations}",
      journal = {\apj},
         year = 2011,
        month = feb,
       volume = {728},
       number = {2},
          eid = {82},
        pages = {82},
          doi = {10.1088/0004-637X/728/2/82},
archivePrefix = {arXiv},
       eprint = {1010.5209},
 primaryClass = {astro-ph.CO},
       adsurl = {https://ui.adsabs.harvard.edu/abs/2011ApJ...728...82M}
}

@ARTICLE{Markevitch2000,
       author = {{Markevitch}, M. and {Ponman}, T.~J. and {Nulsen}, P.~E.~J. and {Bautz}, M.~W. and {Burke}, D.~J. and {David}, L.~P. and {Davis}, D. and {Donnelly}, R.~H. and {Forman}, W.~R. and {Jones}, C. and {Kaastra}, J. and {Kellogg}, E. and {Kim}, D. -W. and {Kolodziejczak}, J. and {Mazzotta}, P. and {Pagliaro}, A. and {Patel}, S. and {Van Speybroeck}, L. and {Vikhlinin}, A. and {Vrtilek}, J. and {Wise}, M. and {Zhao}, P.},
        title = "{Chandra Observation of Abell 2142: Survival of Dense Subcluster Cores in a Merger}",
      journal = {\apj},
         year = 2000,
        month = oct,
       volume = {541},
       number = {2},
        pages = {542-549},
          doi = {10.1086/309470},
archivePrefix = {arXiv},
       eprint = {astro-ph/0001269},
 primaryClass = {astro-ph},
       adsurl = {https://ui.adsabs.harvard.edu/abs/2000ApJ...541..542M}
}

@ARTICLE{Markevitch2001,
       author = {{Markevitch}, M. and {Vikhlinin}, A. and {Mazzotta}, P.},
        title = "{Nonhydrostatic Gas in the Core of the Relaxed Galaxy Cluster A1795}",
      journal = {\apjl},
         year = 2001,
        month = dec,
       volume = {562},
       number = {2},
        pages = {L153-L156},
          doi = {10.1086/337973},
archivePrefix = {arXiv},
       eprint = {astro-ph/0108520},
 primaryClass = {astro-ph},
       adsurl = {https://ui.adsabs.harvard.edu/abs/2001ApJ...562L.153M}
}

@ARTICLE{Markevitch2002,
       author = {{Markevitch}, M. and {Gonzalez}, A.~H. and {David}, L. and {Vikhlinin}, A. and {Murray}, S. and {Forman}, W. and {Jones}, C. and {Tucker}, W.},
        title = "{A Textbook Example of a Bow Shock in the Merging Galaxy Cluster 1E 0657-56}",
      journal = {\apjl},
         year = 2002,
        month = mar,
       volume = {567},
       number = {1},
        pages = {L27-L31},
          doi = {10.1086/339619},
archivePrefix = {arXiv},
       eprint = {astro-ph/0110468},
 primaryClass = {astro-ph},
       adsurl = {https://ui.adsabs.harvard.edu/abs/2002ApJ...567L..27M}
}

@INPROCEEDINGS{Markevitch2003,
       author = {{Markevitch}, M. and {Vikhlinin}, A. and {Forman}, W.~R.},
        title = "{A High Resolution Picture of the Intracluster Gas}",
    booktitle = {Matter and Energy in Clusters of Galaxies},
         year = 2003,
       editor = {{Bowyer}, Stuart and {Hwang}, Chorng-Yuan},
       series = {Astronomical Society of the Pacific Conference Series},
       volume = {301},
        month = jan,
        pages = {37},
          doi = {10.48550/arXiv.astro-ph/0208208},
archivePrefix = {arXiv},
       eprint = {astro-ph/0208208},
 primaryClass = {astro-ph},
       adsurl = {https://ui.adsabs.harvard.edu/abs/2003ASPC..301...37M}
}

@ARTICLE{Markevitch2007,
       author = {{Markevitch}, Maxim and {Vikhlinin}, Alexey},
        title = "{Shocks and cold fronts in galaxy clusters}",
      journal = {\physrep},
         year = 2007,
        month = may,
       volume = {443},
       number = {1},
        pages = {1-53},
          doi = {10.1016/j.physrep.2007.01.001},
archivePrefix = {arXiv},
       eprint = {astro-ph/0701821},
 primaryClass = {astro-ph},
       adsurl = {https://ui.adsabs.harvard.edu/abs/2007PhR...443....1M}
}

@ARTICLE{Mazzotta2008,
       author = {{Mazzotta}, Pasquale and {Giacintucci}, Simona},
        title = "{Do Radio Core-Halos and Cold Fronts in Non-Major-Merging Clusters Originate from the Same Gas Sloshing?}",
      journal = {\apjl},
         year = 2008,
        month = mar,
       volume = {675},
       number = {1},
        pages = {L9},
          doi = {10.1086/529433},
archivePrefix = {arXiv},
       eprint = {0801.1905},
 primaryClass = {astro-ph},
       adsurl = {https://ui.adsabs.harvard.edu/abs/2008ApJ...675L...9M}
}

@ARTICLE{Miniati2000,
       author = {{Miniati}, Francesco and {Ryu}, Dongsu and {Kang}, Hyesung and {Jones}, T.~W. and {Cen}, Renyue and {Ostriker}, Jeremiah P.},
        title = "{Properties of Cosmic Shock Waves in Large-Scale Structure Formation}",
      journal = {\apj},
         year = 2000,
        month = oct,
       volume = {542},
       number = {2},
        pages = {608-621},
          doi = {10.1086/317027},
archivePrefix = {arXiv},
       eprint = {astro-ph/0005444},
 primaryClass = {astro-ph},
       adsurl = {https://ui.adsabs.harvard.edu/abs/2000ApJ...542..608M}
}

@ARTICLE{Mirakhor2023,
       author = {{Mirakhor}, M.~S. and {Walker}, S.~A. and {Sundquist}, M. and {Chandra}, D.},
        title = "{Two large-scale sloshing cold fronts in the outskirts of the galaxy cluster Abell 3558}",
      journal = {\mnras},
         year = 2023,
        month = nov,
       volume = {526},
       number = {1},
        pages = {L124-L128},
          doi = {10.1093/mnrasl/slad129},
archivePrefix = {arXiv},
       eprint = {2308.16222},
 primaryClass = {astro-ph.CO},
       adsurl = {https://ui.adsabs.harvard.edu/abs/2023MNRAS.526L.124M}
}

@ARTICLE{Nulsen2005,
       author = {{Nulsen}, P.~E.~J. and {McNamara}, B.~R. and {Wise}, M.~W. and {David}, L.~P.},
        title = "{The Cluster-Scale AGN Outburst in Hydra A}",
      journal = {\apj},
         year = 2005,
        month = aug,
       volume = {628},
       number = {2},
        pages = {629-636},
          doi = {10.1086/430845},
archivePrefix = {arXiv},
       eprint = {astro-ph/0408315},
 primaryClass = {astro-ph},
       adsurl = {https://ui.adsabs.harvard.edu/abs/2005ApJ...628..629N}
}

@ARTICLE{Oppizzi2023,
       author = {{Oppizzi}, F. and {De Luca}, F. and {Bourdin}, H. and {Mazzotta}, P. and {Ettori}, S. and {Gastaldello}, F. and {Kay}, S. and {Lovisari}, L. and {Maughan}, B.~J. and {Pointecouteau}, E. and {Pratt}, G.~W. and {Rossetti}, M. and {Sayers}, J. and {Sereno}, M.},
        title = "{CHEX-MATE: Pressure profiles of six galaxy clusters as seen by SPT and Planck}",
      journal = {\aap},
         year = 2023,
        month = apr,
       volume = {672},
          eid = {A156},
        pages = {A156},
          doi = {10.1051/0004-6361/202245012},
archivePrefix = {arXiv},
       eprint = {2209.09601},
 primaryClass = {astro-ph.CO},
       adsurl = {https://ui.adsabs.harvard.edu/abs/2023A\&A...672A.156O}
}

@ARTICLE{Owers2009,
       author = {{Owers}, Matt S. and {Nulsen}, Paul E.~J. and {Couch}, Warrick J. and {Markevitch}, Maxim and {Poole}, Gregory B.},
        title = "{Abell 1201: The Anatomy of a Cold Front Cluster from Combined Optical and X-Ray Data}",
      journal = {\apj},
         year = 2009,
        month = feb,
       volume = {692},
       number = {1},
        pages = {702-722},
          doi = {10.1088/0004-637X/692/1/702},
archivePrefix = {arXiv},
       eprint = {0810.4650},
 primaryClass = {astro-ph},
       adsurl = {https://ui.adsabs.harvard.edu/abs/2009ApJ...692..702O}
}

@article{ Planck205,
	author = {{Planck Collaboration} and {Ade, P. A. R.} and {Aghanim, N.} and {Arnaud, M.} and {Ashdown, M.} and {Aumont, J.} and {Baccigalupi, C.} and {Banday, A. J.} and {Barreiro, R. B.} and {Barrena, R.} and {Bartlett, J. G.} and {Bartolo, N.} and {Battaner, E.} and {Battye, R.} and {Benabed, K.} and {Benoît, A.} and {Benoit-Lévy, A.} and {Bernard, J.-P.} and {Bersanelli, M.} and {Bielewicz, P.} and {Bikmaev, I.} and {Böhringer, H.} and {Bonaldi, A.} and {Bonavera, L.} and {Bond, J. R.} and {Borrill, J.} and {Bouchet, F. R.} and {Bucher, M.} and {Burenin, R.} and {Burigana, C.} and {Butler, R. C.} and {Calabrese, E.} and {Cardoso, J.-F.} and {Carvalho, P.} and {Catalano, A.} and {Challinor, A.} and {Chamballu, A.} and {Chary, R.-R.} and {Chiang, H. C.} and {Chon, G.} and {Christensen, P. R.} and {Clements, D. L.} and {Colombi, S.} and {Colombo, L. P. L.} and {Combet, C.} and {Comis, B.} and {Couchot, F.} and {Coulais, A.} and {Crill, B. P.} and {Curto, A.} and {Cuttaia, F.} and {Dahle, H.} and {Danese, L.} and {Davies, R. D.} and {Davis, R. J.} and {de Bernardis, P.} and {de Rosa, A.} and {de Zotti, G.} and {Delabrouille, J.} and {Désert, F.-X.} and {Dickinson, C.} and {Diego, J. M.} and {Dolag, K.} and {Dole, H.} and {Donzelli, S.} and {Doré, O.} and {Douspis, M.} and {Ducout, A.} and {Dupac, X.} and {Efstathiou, G.} and {Eisenhardt, P. R. M.} and {Elsner, F.} and {Enßlin, T. A.} and {Eriksen, H. K.} and {Falgarone, E.} and {Fergusson, J.} and {Feroz, F.} and {Ferragamo, A.} and {Finelli, F.} and {Forni, O.} and {Frailis, M.} and {Fraisse, A. A.} and {Franceschi, E.} and {Frejsel, A.} and {Galeotta, S.} and {Galli, S.} and {Ganga, K.} and {Génova-Santos, R. T.} and {Giard, M.} and {Giraud-Héraud, Y.} and {Gjerløw, E.} and {González-Nuevo, J.} and {Górski, K. M.} and {Grainge, K. J. B.} and {Gratton, S.} and {Gregorio, A.} and {Gruppuso, A.} and {Gudmundsson, J. E.} and {Hansen, F. K.} and {Hanson, D.} and {Harrison, D. L.} and {Hempel, A.} and {Henrot-Versillé, S.} and {Hernández-Monteagudo, C.} and {Herranz, D.} and {Hildebrandt, S. R.} and {Hivon, E.} and {Hobson, M.} and {Holmes, W. A.} and {Hornstrup, A.} and {Hovest, W.} and {Huffenberger, K. M.} and {Hurier, G.} and {Jaffe, A. H.} and {Jaffe, T. R.} and {Jin, T.} and {Jones, W. C.} and {Juvela, M.} and {Keihänen, E.} and {Keskitalo, R.} and {Khamitov, I.} and {Kisner, T. S.} and {Kneissl, R.} and {Knoche, J.} and {Kunz, M.} and {Kurki-Suonio, H.} and {Lagache, G.} and {Lamarre, J.-M.} and {Lasenby, A.} and {Lattanzi, M.} and {Lawrence, C. R.} and {Leonardi, R.} and {Lesgourgues, J.} and {Levrier, F.} and {Liguori, M.} and {Lilje, P. B.} and {Linden-Vørnle, M.} and {López-Caniego, M.} and {Lubin, P. M.} and {Macías-Pérez, J. F.} and {Maggio, G.} and {Maino, D.} and {Mak, D. S. Y.} and {Mandolesi, N.} and {Mangilli, A.} and {Martin, P. G.} and {Martínez-González, E.} and {Masi, S.} and {Matarrese, S.} and {Mazzotta, P.} and {McGehee, P.} and {Mei, S.} and {Melchiorri, A.} and {Melin, J.-B.} and {Mendes, L.} and {Mennella, A.} and {Migliaccio, M.} and {Mitra, S.} and {Miville-Deschênes, M.-A.} and {Moneti, A.} and {Montier, L.} and {Morgante, G.} and {Mortlock, D.} and {Moss, A.} and {Munshi, D.} and {Murphy, J. A.} and {Naselsky, P.} and {Nastasi, A.} and {Nati, F.} and {Natoli, P.} and {Netterfield, C. B.} and {Nørgaard-Nielsen, H. U.} and {Noviello, F.} and {Novikov, D.} and {Novikov, I.} and {Olamaie, M.} and {Oxborrow, C. A.} and {Paci, F.} and {Pagano, L.} and {Pajot, F.} and {Paoletti, D.} and {Pasian, F.} and {Patanchon, G.} and {Pearson, T. J.} and {Perdereau, O.} and {Perotto, L.} and {Perrott, Y. C.} and {Perrotta, F.} and {Pettorino, V.} and {Piacentini, F.} and {Piat, M.} and {Pierpaoli, E.} and {Pietrobon, D.} and {Plaszczynski, S.} and {Pointecouteau, E.} and {Polenta, G.} and {Pratt, G. W.} and {Prézeau, G.} and {Prunet, S.} and {Puget, J.-L.} and {Rachen, J. P.} and {Reach, W. T.} and {Rebolo, R.} and {Reinecke, M.} and {Remazeilles, M.} and {Renault, C.} and {Renzi, A.} and {Ristorcelli, I.} and {Rocha, G.} and {Rosset, C.} and {Rossetti, M.} and {Roudier, G.} and {Rozo, E.} and {Rubiño-Martín, J. A.} and {Rumsey, C.} and {Rusholme, B.} and {Rykoff, E. S.} and {Sandri, M.} and {Santos, D.} and {Saunders, R. D. E.} and {Savelainen, M.} and {Savini, G.} and {Schammel, M. P.} and {Scott, D.} and {Seiffert, M. D.} and {Shellard, E. P. S.} and {Shimwell, T. W.} and {Spencer, L. D.} and {Stanford, S. A.} and {Stern, D.} and {Stolyarov, V.} and {Stompor, R.} and {Streblyanska, A.} and {Sudiwala, R.} and {Sunyaev, R.} and {Sutton, D.} and {Suur-Uski, A.-S.} and {Sygnet, J.-F.} and {Tauber, J. A.} and {Terenzi, L.} and {Toffolatti, L.} and {Tomasi, M.} and {Tramonte, D.} and {Tristram, M.} and {Tucci, M.} and {Tuovinen, J.} and {Umana, G.} and {Valenziano, L.} and {Valiviita, J.} and {Van Tent, B.} and {Vielva, P.} and {Villa, F.} and {Wade, L. A.} and {Wandelt, B. D.} and {Wehus, I. K.} and {White, S. D. M.} and {Wright, E. L.} and {Yvon, D.} and {Zacchei, A.} and {Zonca, A.}},
	title = {Planck 2015 results - XXVII. The second Planck catalogue of Sunyaev-Zeldovich sources},
	DOI= "10.1051/0004-6361/201525823",
	url= "https://doi.org/10.1051/0004-6361/201525823",
	journal = {A\&A},
	year = 2016,
	volume = 594,
	pages = "A27",
}

@ARTICLE{Quilis2001,
       author = {{Quilis}, Vicent and {Bower}, Richard G. and {Balogh}, Michael L.},
        title = "{Bubbles, feedback and the intracluster medium: three-dimensional hydrodynamic simulations}",
      journal = {\mnras},
         year = 2001,
        month = dec,
       volume = {328},
       number = {4},
        pages = {1091-1097},
          doi = {10.1046/j.1365-8711.2001.04927.x},
archivePrefix = {arXiv},
       eprint = {astro-ph/0109022},
 primaryClass = {astro-ph},
       adsurl = {https://ui.adsabs.harvard.edu/abs/2001MNRAS.328.1091Q}
}

@ARTICLE{Roediger2011,
       author = {{Roediger}, E. and {Br{\"u}ggen}, M. and {Simionescu}, A. and {B{\"o}hringer}, H. and {Churazov}, E. and {Forman}, W.~R.},
        title = "{Gas sloshing, cold front formation and metal redistribution: the Virgo cluster as a quantitative test case}",
      journal = {\mnras},
         year = 2011,
        month = may,
       volume = {413},
       number = {3},
        pages = {2057-2077},
          doi = {10.1111/j.1365-2966.2011.18279.x},
archivePrefix = {arXiv},
       eprint = {1007.4209},
 primaryClass = {astro-ph.CO},
       adsurl = {https://ui.adsabs.harvard.edu/abs/2011MNRAS.413.2057R}
}

@ARTICLE{Roediger2012,
       author = {{Roediger}, E. and {Lovisari}, L. and {Dupke}, R. and {Ghizzardi}, S. and {Br{\"u}ggen}, M. and {Kraft}, R.~P. and {Machacek}, M.~E.},
        title = "{Gas sloshing, cold fronts, Kelvin-Helmholtz instabilities and the merger history of the cluster of galaxies Abell 496}",
      journal = {\mnras},
         year = 2012,
        month = mar,
       volume = {420},
       number = {4},
        pages = {3632-3648},
          doi = {10.1111/j.1365-2966.2011.20287.x},
archivePrefix = {arXiv},
       eprint = {1112.1407},
 primaryClass = {astro-ph.CO},
       adsurl = {https://ui.adsabs.harvard.edu/abs/2012MNRAS.420.3632R}
}

@ARTICLE{Rossetti2013,
       author = {{Rossetti}, M. and {Eckert}, D. and {De Grandi}, S. and {Gastaldello}, F. and {Ghizzardi}, S. and {Roediger}, E. and {Molendi}, S.},
        title = "{Abell 2142 at large scales: An extreme case for sloshing?}",
      journal = {\aap},
         year = 2013,
        month = aug,
       volume = {556},
          eid = {A44},
        pages = {A44},
          doi = {10.1051/0004-6361/201321319},
archivePrefix = {arXiv},
       eprint = {1305.2420},
 primaryClass = {astro-ph.CO},
       adsurl = {https://ui.adsabs.harvard.edu/abs/2013A\&A...556A..44R}
}

@ARTICLE{Sanders2016,
       author = {{Sanders}, J.~S. and {Fabian}, A.~C. and {Russell}, H.~R. and {Walker}, S.~A. and {Blundell}, K.~M.},
        title = "{Detecting edges in the X-ray surface brightness of galaxy clusters}",
      journal = {\mnras},
         year = 2016,
        month = aug,
       volume = {460},
       number = {2},
        pages = {1898-1911},
          doi = {10.1093/mnras/stw1119},
archivePrefix = {arXiv},
       eprint = {1605.02911},
 primaryClass = {astro-ph.CO},
       adsurl = {https://ui.adsabs.harvard.edu/abs/2016MNRAS.460.1898S}
}

@ARTICLE{Santra2024,
       author = {{Santra}, R. and {Kale}, R. and {Giacintucci}, S. and {Markevitch}, M. and {De Luca}, F. and {Bourdin}, H. and {Venturi}, T. and {Dallacasa}, D. and {Cassano}, R. and {Brunetti}, G. and {Buch}, K.~D.},
        title = "{A Deep uGMRT View of the Ultra-steep Spectrum Radio Halo in A521}",
      journal = {\apj},
         year = 2024,
        month = feb,
       volume = {962},
       number = {1},
          eid = {40},
        pages = {40},
          doi = {10.3847/1538-4357/ad1190},
archivePrefix = {arXiv},
       eprint = {2311.09717},
 primaryClass = {astro-ph.CO},
       adsurl = {https://ui.adsabs.harvard.edu/abs/2024ApJ...962...40S}
}

@ARTICLE{Shimwell2015,
       author = {{Shimwell}, Timothy W. and {Markevitch}, Maxim and {Brown}, Shea and {Feretti}, Luigina and {Gaensler}, B.~M. and {Johnston-Hollitt}, M. and {Lage}, Craig and {Srinivasan}, Raghav},
        title = "{Another shock for the Bullet cluster, and the source of seed electrons for radio relics}",
      journal = {\mnras},
         year = 2015,
        month = may,
       volume = {449},
       number = {2},
        pages = {1486-1494},
          doi = {10.1093/mnras/stv334},
archivePrefix = {arXiv},
       eprint = {1502.01064},
 primaryClass = {astro-ph.HE},
       adsurl = {https://ui.adsabs.harvard.edu/abs/2015MNRAS.449.1486S}
}

@ARTICLE{Simionescu2012,
       author = {{Simionescu}, A. and {Werner}, N. and {Urban}, O. and {Allen}, S.~W. and {Fabian}, A.~C. and {Sanders}, J.~S. and {Mantz}, A. and {Nulsen}, P.~E.~J. and {Takei}, Y.},
        title = "{Large-scale Motions in the Perseus Galaxy Cluster}",
      journal = {\apj},
         year = 2012,
        month = oct,
       volume = {757},
       number = {2},
          eid = {182},
        pages = {182},
          doi = {10.1088/0004-637X/757/2/182},
archivePrefix = {arXiv},
       eprint = {1208.2990},
 primaryClass = {astro-ph.CO},
       adsurl = {https://ui.adsabs.harvard.edu/abs/2012ApJ...757..182S}
}

@article{Sobel1973,
author = {Sobel, Irwin and Feldman, Gary},
year = {1973},
month = {01},
pages = {271-272},
title = {A 3×3 isotropic gradient operator for image processing},
journal = {Pattern Classification and Scene Analysis}
}

@ARTICLE{Snowden2008,
       author = {{Snowden}, S.~L. and {Mushotzky}, R.~F. and {Kuntz}, K.~D. and {Davis}, D.~S.},
        title = "{A catalog of galaxy clusters observed by XMM-Newton}",
      journal = {\aap},
         year = 2008,
        month = feb,
       volume = {478},
       number = {2},
        pages = {615-658},
          doi = {10.1051/0004-6361:20077930},
archivePrefix = {arXiv},
       eprint = {0710.2241},
 primaryClass = {astro-ph},
       adsurl = {https://ui.adsabs.harvard.edu/abs/2008A\&A...478..615S}
}

@article{xcop,
        Adsurl = {http://adsabs.harvard.edu/abs/2017AN....338..293E},
        Archiveprefix = {arXiv},
        Author = {{Eckert}, D. and {Ettori}, S. and {Pointecouteau}, E. and {Molendi}, S. and {Paltani}, S. and {Tchernin}, C.},
        Doi = {10.1002/asna.201713345},
        Eprint = {1611.05051},
        Journal = {Astronomische Nachrichten},
        Month = mar,
        Pages = {293-298},
        Title = {{The XMM cluster outskirts project (X-COP)}},
        Volume = 338,
        Year = 2017}

@ARTICLE{Ubertosi2023,
       author = {{Ubertosi}, F. and {Gitti}, M. and {Brighenti}, F. and {McDonald}, M. and {Nulsen}, P. and {Donahue}, M. and {Brunetti}, G. and {Randall}, S. and {Gaspari}, M. and {Ettori}, S. and {Calzadilla}, M. and {Ignesti}, A. and {Feretti}, L. and {Blanton}, E.~L.},
        title = "{Multiple Shock Fronts in RBS 797: The Chandra Window on Shock Heating in Galaxy Clusters}",
      journal = {\apj},
         year = 2023,
        month = feb,
       volume = {944},
       number = {2},
          eid = {216},
        pages = {216},
          doi = {10.3847/1538-4357/acacf9},
archivePrefix = {arXiv},
       eprint = {2212.10581},
 primaryClass = {astro-ph.GA},
       adsurl = {https://ui.adsabs.harvard.edu/abs/2023ApJ...944..216U}
}

@ARTICLE{Ueda2019,
       author = {{Ueda}, Shutaro and {Ichinohe}, Yuto and {Kitayama}, Tetsu and {Umetsu}, Keiichi},
        title = "{Line-of-Sight Gas Sloshing in the Cool Core of Abell 907}",
      journal = {\apj},
         year = 2019,
        month = feb,
       volume = {871},
       number = {2},
          eid = {207},
        pages = {207},
          doi = {10.3847/1538-4357/aafa19},
archivePrefix = {arXiv},
       eprint = {1812.07835},
 primaryClass = {astro-ph.HE},
       adsurl = {https://ui.adsabs.harvard.edu/abs/2019ApJ...871..207U}
}

@ARTICLE{vanweeren2019,
       author = {{van Weeren}, R.~J. and {de Gasperin}, F. and {Akamatsu}, H. and {Br{\"u}ggen}, M. and {Feretti}, L. and {Kang}, H. and {Stroe}, A. and {Zandanel}, F.},
        title = "{Diffuse Radio Emission from Galaxy Clusters}",
      journal = {\ssr},
         year = 2019,
        month = feb,
       volume = {215},
       number = {1},
          eid = {16},
        pages = {16},
          doi = {10.1007/s11214-019-0584-z},
archivePrefix = {arXiv},
       eprint = {1901.04496},
 primaryClass = {astro-ph.HE},
       adsurl = {https://ui.adsabs.harvard.edu/abs/2019SSRv..215...16V}
}

@ARTICLE{Vikhlinin2001a,
       author = {{Vikhlinin}, A. and {Markevitch}, M. and {Murray}, S.~S.},
        title = "{A Moving Cold Front in the Intergalactic Medium of A3667}",
      journal = {\apj},
         year = 2001,
        month = apr,
       volume = {551},
       number = {1},
        pages = {160-171},
          doi = {10.1086/320078},
archivePrefix = {arXiv},
       eprint = {astro-ph/0008496},
 primaryClass = {astro-ph},
       adsurl = {https://ui.adsabs.harvard.edu/abs/2001ApJ...551..160V}
}

@ARTICLE{Vikhlinin2001b,
       author = {{Vikhlinin}, A. and {Markevitch}, M. and {Forman}, W. and {Jones}, C.},
        title = "{Zooming in on the Coma Cluster with Chandra: Compressed Warm Gas in the Brightest Cluster Galaxies}",
      journal = {\apjl},
         year = 2001,
        month = jul,
       volume = {555},
       number = {2},
        pages = {L87-L90},
          doi = {10.1086/323181},
archivePrefix = {arXiv},
       eprint = {astro-ph/0102483},
 primaryClass = {astro-ph},
       adsurl = {https://ui.adsabs.harvard.edu/abs/2001ApJ...555L..87V}
}

@ARTICLE{Vikhlinin2002,
       author = {{Vikhlinin}, A.~A. and {Markevitch}, M.~L.},
        title = "{A Cold Front in the Galaxy Cluster A3667: Hydrodynamics, Heat Conduction and Magnetic Field in the Intergalactic Medium}",
      journal = {Astronomy Letters},
         year = 2002,
        month = aug,
       volume = {28},
        pages = {495-508},
          doi = {10.1134/1.1499173},
archivePrefix = {arXiv},
       eprint = {astro-ph/0209551},
 primaryClass = {astro-ph},
       adsurl = {https://ui.adsabs.harvard.edu/abs/2002AstL...28..495V}
}

@ARTICLE{Walker2022,
       author = {{Walker}, S.~A. and {Mirakhor}, M.~S. and {ZuHone}, J. and {Sanders}, J.~S. and {Fabian}, A.~C. and {Diwanji}, P.},
        title = "{Is There an Enormous Cold Front at the Virial Radius of the Perseus Cluster?}",
      journal = {\apj},
         year = 2022,
        month = apr,
       volume = {929},
       number = {1},
          eid = {37},
        pages = {37},
          doi = {10.3847/1538-4357/ac5894},
archivePrefix = {arXiv},
       eprint = {2006.14043},
 primaryClass = {astro-ph.HE},
       adsurl = {https://ui.adsabs.harvard.edu/abs/2022ApJ...929...37W}
}

@ARTICLE{Wang2018,
       author = {{Wang}, Qian H.~S. and {Markevitch}, Maxim},
        title = "{A Deep X-Ray Look at Abell 2142{\textemdash}Viscosity Constraints From Kelvin-Helmholtz Eddies, a Displaced Cool Peak That Makes a Warm Core, and A Possible Plasma Depletion Layer}",
      journal = {\apj},
         year = 2018,
        month = nov,
       volume = {868},
       number = {1},
          eid = {45},
        pages = {45},
          doi = {10.3847/1538-4357/aae921},
archivePrefix = {arXiv},
       eprint = {1810.02813},
 primaryClass = {astro-ph.HE},
       adsurl = {https://ui.adsabs.harvard.edu/abs/2018ApJ...868...45W}
}

@ARTICLE{Zhang2020,
       author = {{Zhang}, Congyao and {Churazov}, Eugene and {Dolag}, Klaus and {Forman}, William R. and {Zhuravleva}, Irina},
        title = "{Collision of merger and accretion shocks: formation of Mpc-scale contact discontinuity in the Perseus cluster}",
      journal = {\mnras},
         year = 2020,
        month = oct,
       volume = {498},
       number = {1},
        pages = {L130-L134},
          doi = {10.1093/mnrasl/slaa147},
archivePrefix = {arXiv},
       eprint = {2007.02551},
 primaryClass = {astro-ph.HE},
       adsurl = {https://ui.adsabs.harvard.edu/abs/2020MNRAS.498L.130Z}
}

@ARTICLE{ZuHone2013b,
       author = {{ZuHone}, J.~A. and {Markevitch}, M. and {Ruszkowski}, M. and {Lee}, D.},
        title = "{Cold Fronts and Gas Sloshing in Galaxy Clusters with Anisotropic Thermal Conduction}",
      journal = {\apj},
         year = 2013,
        month = jan,
       volume = {762},
       number = {2},
          eid = {69},
        pages = {69},
          doi = {10.1088/0004-637X/762/2/69},
archivePrefix = {arXiv},
       eprint = {1204.6005},
 primaryClass = {astro-ph.CO},
       adsurl = {https://ui.adsabs.harvard.edu/abs/2013ApJ...762...69Z}
}

@ARTICLE{zuhone2016,
       author = {{Zuhone}, John A. and {Roediger}, E.},
        title = "{Cold fronts: probes of plasma astrophysics in galaxy clusters}",
      journal = {Journal of Plasma Physics},
         year = 2016,
        month = jun,
       volume = {82},
       number = {3},
          eid = {535820301},
        pages = {535820301},
          doi = {10.1017/S0022377816000544},
archivePrefix = {arXiv},
       eprint = {1603.08882},
 primaryClass = {astro-ph.HE},
       adsurl = {https://ui.adsabs.harvard.edu/abs/2016JPlPh..82c5301Z}
}

\end{document}